\documentclass[pdfpagelabels=false,colorlinks,citecolor=blue,useAMS,usenatbib]{mnras}

\usepackage{epsfig,graphicx,mathptmx}
\usepackage{times}
\usepackage{natbib}
\usepackage{amsmath}
\usepackage{amssymb}	% Extra maths symbols
\usepackage{graphicx}	% Including figure files
\usepackage[dvipsnames]{xcolor}
\usepackage{soul}

\newcommand{\mone}{$^{-1}$}
\newcommand{\sqr}{$^2$}
\newcommand{\cub}{$^3$}
\newcommand{\bpar}{$b$-parameter}
\newcommand{\lcdm}{$\Lambda$CDM}

\newcommand{\muk}{$\umu$K}
\newcommand{\omegam}{$\Omega_{\rm m}$}
\newcommand{\omegacdm}{$\Omega_{\rm CDM}$}
\newcommand{\omegal}{$\Omega_\Lambda$}
\newcommand{\omegab}{$\Omega_{\rm b}$}

\newcommand{\sigmae}{$\sigma_8$}
\newcommand{\hzero}{$H_0$}
\newcommand{\fnu}{$f_\nu$}
\newcommand{\frz}{$\overline{f}_{\rm R,0}$}
\newcommand{\frza}{$|\overline{f}_{\rm R,0}|$}
\newcommand{\frztnz}{\frz$=(-10^{-6},-10^{-5},-10^{-4})$}
\newcommand{\frzt}{\frz$=(0,-10^{-6},-10^{-5},-10^{-4})$}

\newcommand{\zre}{$z_{\rm re}$}

\newcommand{\gadgetiii}{\textsc{gadget-iii}}
\newcommand{\msun}{$M_{\sun}$}
\newcommand{\hmone}{$\,h^{-1}$}
\newcommand{\dksz}{$\mathcal{D}^{\rm kSZ}_{\ell}$}
\newcommand{\dkszt}{$\mathcal{D}^{\rm kSZ}_{3000}$}
\newcommand{\dksztp}{$\mathcal{D}^{\rm kSZ,patchy}_{3000}$}

\newcommand{\mggadg}{\textsc{mg-gadget}}
\DeclareMathAlphabet{\mathcal}{OMS}{cmsy}{m}{n}

\def\prd{Phys. Rev. D}
\def\aap{A\&A}
\def\apj{ApJ}

\def\apjl{ApJ}
\def\mnras{MNRAS}

\def\aj{AJ}

\def\nat{Nat}

\def\apss{Ap\&SS}      % Astrophysics and Space Science
\def\apjs{ApJS}
\def\jcap{JCAP}        % Journal of Cosmology and Astro-Particle Physics

\title[The kSZ effect and f(R)]
{The kinematic Sunyaev--Zel'dovich effect of the large-scale structure~(II): the effect 
of modified gravity}
\author[M. Roncarelli et al.]
{M. Roncarelli$^{1,2}$\thanks{E-mail: \href{mailto:mauro.roncarelli@unibo.it}{mauro.roncarelli@unibo.it}},
M. Baldi$^{1,2,3}$ and F. Villaescusa-Navarro$^{4,5,6}$ \\
\\
$^1$Dipartimento di Fisica e Astronomia, Universit\`a di Bologna, viale Berti Pichat 
6/2, I-40127 Bologna, Italy \\
$^2$Istituto Nazionale di Astrofisica (INAF) -- Osservatorio di Astrofisica e Scienza 
dello Spazio (OAS), via Gobetti 93/3, I-40127 Bologna, Italy \\
$^3$Istituto Nazionale di Fisica Nucleare (INFN) - Sezione di Bologna, viale Berti 
Pichat 6/2, I-40127 Bologna, Italy \\
$^4$Center for Computational Astrophysics, 160 Fifth Avenue, New York, NY 10010, USA \\
$^5$Istituto Nazionale di Astrofisica (INAF) -- Osservatorio Astronomico di Trieste, via 
Tiepolo 11, I-34131 Trieste, Italy \\
$^6$Istituto Nazionale di Fisica Nucleare (INFN) -- Sezione di Trieste, via Valerio 2, 
I-34127 Trieste, Italy
}

\begin{document}

\date{Accepted  ??. Received ??; in original form ??}

\pagerange{\pageref{firstpage}--\pageref{lastpage}} \pubyear{2018}

\maketitle

\label{firstpage}

\begin{abstract}
  The key to understand the nature of dark energy relies in our ability to probe the 
  distant Universe. In this framework, the recent detection of the kinematic 
  Sunyaev-Zel'dovich (kSZ) effect signature in the cosmic microwave background obtained 
  with the South Pole Telescope (SPT) is extremely useful since this observable is 
  sensitive to the high-redshift diffuse plasma. We analyse a set of cosmological 
  hydrodynamical simulation with 4 different realisations of a Hu \& Sawicki $f(R)$ 
  gravity model, parametrised by the values of \frzt, to compute the properties of the 
  kSZ effect due to the ionized Universe and how they depend on \frz\ and on the redshift 
  of reionization, \zre. In the standard General Relativity limit (\frz=0) we obtain an 
  amplitude of the kSZ power spectrum of \dkszt$=4.1\,$\muk\sqr\ (\zre=8.8), close to the 
  $+1\sigma$ limit of the \dkszt$=(2.9\pm1.3)\,$\muk\sqr\ measurement by SPT. This 
  corresponds to an upper limit on the kSZ contribute from patchy reionization of 
  \dksztp$<0.9\,$\muk\sqr\ (95 per cent confidence level). Modified gravity boosts the 
  kSZ signal by about 3, 12 and 50 per cent for \frztnz, respectively, with almost no 
  dependence on the angular scale. This means that with modified gravity the limits on 
  patchy reionization shrink significantly: for \frz$=-10^{-5}$ we obtain 
  \dksztp$<0.4\,$\muk\sqr. Finally, we provide an analytical formula for the scaling of 
  the kSZ power spectrum with \zre\ and \frz\ at different multipoles: at $\ell=3000$ we 
  obtain \dkszt$\propto$
  \zre$^{0.24}\left(1+\sqrt{\left|\overline{f}_{\rm R,0}\right|}\right)^{41}$.
\end{abstract}


%%%%%%%%%%%%%%%%%%%%%%%%%%%%%%%%%%%%%%%%%%%%%%%%%%%%%%%%%%%%%%%%%%%%%%%%%%%%%%%%%%%%%%%%%%
%%%%%%%%%%%%%%%%%%%%%%%%%%%%%%%%%%%%%% INTRODUCTION %%%%%%%%%%%%%%%%%%%%%%%%%%%%%%%%%%%%%%
%%%%%%%%%%%%%%%%%%%%%%%%%%%%%%%%%%%%%%%%%%%%%%%%%%%%%%%%%%%%%%%%%%%%%%%%%%%%%%%%%%%%%%%%%%

\section{INTRODUCTION}
\label{s:intro}

The origin of the accelerated expansion of the Universe has become a longstanding problem 
in cosmology and in theoretical physics in general: we are about to mark the {\em 20th} 
anniversary of the first detection \citep[by means of the dimming of distant supernovae]
[]{Riess_etal_1998,Schmidt_etal_1998,Perlmutter_etal_1999} of such pivotal discovery that 
shaped the development of a whole field of research, and that provided support and 
motivations for the design and funding of several challenging and costly observational 
enterprises to survey large portions of the sky, both from the ground (DES, LSST, HetDEX) 
and from space \citep[Euclid, see][]{Euclid-r}. 

On the theoretical side, the past two decades have been characterised by widespread 
efforts to provide a more solid and natural framework for the accelerated expansion with 
respect to the highly fine-tuned (and yet most economic) option of the cosmological 
constant $\Lambda $ that characterises the current standard cosmological model. Possible 
alternative scenarios have attempted to invoke the slow roll of a light scalar field 
\citep[known as Quintessence][]{Wetterich_1988,Ratra_Peebles_1988} and its possible 
couplings to the matter sector \citep[see e.g.][]{Wetterich_1995,
Amendola_2000,Amendola_Baldi_Wetterich_2008} as 
an explanation for the energy scale of the Dark Energy and for the relatively recent 
onset of the accelerated expansion, known as the {\em fine-tuning} and {\em coincidence} 
problems, respectively. 

An alternative option amounts to considering possible deviations from the standard theory 
of gravity in the form of extensions to General Relativity (GR) that may result in an 
effective weakening of gravity at cosmological scales and late cosmic epochs. Such 
approach, generally known as {\em Modified Gravity}, has been extensively explored over 
the past years \citep[see e.g.][]{Euclid_TWG} allowing to identify and classify a wide 
range of geometric theories of gravity that deviate from standard GR still providing 
healthy equations of motion and sensible cosmological evolutions. As any deviation from 
the predicted behaviour of GR within the Solar System is very tightly constrained by 
local tests of gravity \citep[see e.g.][]{Bertotti_Iess_Tortora_2003,Will_2005}, all such 
theories must rely on some mechanism to recover standard GR within the local environment, 
which go under the general term of {\em screening} \citep[see e.g.][]
{Khoury_Weltman_2004,Vainshtein_1972,Damour_Polyakov_1994,Hinterbichler_Khoury_2010,
Nicolis_Rattazzi_Trincherini_2009}. Unfortunately, when such constraining conditions are 
applied, most modified gravity theories still require a fine-tuned low energy scale to 
reproduce the observed background expansion history, thereby failing to ease the problems 
of the cosmological constant. Nonetheless, these scenarios still provide a theoretically 
consistent framework to test gravity on large scales and constrain possible deviations 
from the standard theory of GR. Besides the Solar System tests, the recent detection of 
the gravitational wave event GW170817 \citep[][]{GW170817} and of its electomagnetic 
counterpart has also severely constrained the landscape of possible modified gravity 
theories \citep[see e.g.][]{Baker_etal_2017,Sakstein_Jain_2017,Creminelli_Vernizzi_2017,
Ezquiaga_Zumalacarregui_2017} by ruling out with a single observation all modified 
gravity modelsfeaturing a non-negligble difference of the propagation velocities of 
electromagntic and gravitational signals. After this selection, only a bunch of modified 
gravity scenarioscan be still considered as viable candidates for an extension of GR at 
cosmological scales \citep[see e.g. Fig.~2 in ][]{Ezquiaga_Zumalacarregui_2017}. \\

The most widely studied example of such models still passing the GW170817 scrutiny is 
given by $f(R)$ gravity \citep[][]{Buchdahl_1970}, where the standard Ricci curvature 
term $R$ in the Einstein-Hilbert Action is extended by an additional function $f(R)$:
\begin{equation}
\label{fRaction}
  S = \int {\rm d}^4x \, \sqrt{-g} \left( \frac{R+f(R)}{16 \pi G} + {\cal L}_m \right).
\end{equation}
In equation~(\ref{fRaction}) $G$ is Newton's gravitational constant, $g$ is the determinant 
of the metric tensor $g_{\mu \nu }$, and ${\cal L}_m$ is the Lagrangian density of all 
matter fields. The quantity $f_R \equiv {\rm d}f(R)/{\rm d}R$ represents a new scalar 
degree of freedom that propagates as the carrier of an additional force. In the 
weak-field and quasi-static limit, this scalar field obeys an independent dynamic 
equation\footnote{For the formulas presented in this section, we work in units where the 
speed of light is set to unity, $c=1$.}
{\citep[see][]{Hu_Sawicki_2007}:}
\begin{equation}
  \nabla^2 f_R = \frac{1}{3}\left(\delta R - 8 \pi G \delta \rho \right) \,,
\label{e:fR_field_eq}
\end{equation}
where $\delta R$ and $\delta \rho$ are the relative perturbations in the scalar curvature 
and matter density, respectively.\\

Different choices for the functional form of $f(R)$ in equation~(\ref{fRaction}) may 
then lead to a large variety of effects on both the background expansion history of the 
Universe and the growth of its density perturbations, giving thus rise to possible 
characteristic observational signatures in the resulting  large-scale structure (LSS) 
formation. Among the many ways to parameterise the variety of possible $f(R)$ forms, the 
most common approach (see Section~\ref{s:fr} for a more detailed discussion) adopts the 
mean value of $f_R$ at the present epoch, \frz, as the key parameter to describe 
deviations from standard GR, and uses it as a reference to define observational 
constraints.

The wide range of diverse phenomenological effects that modified gravity models imprint 
on gravitating systems at different scales -- from the dynamics of the Solar System to 
the expansion of the universe and the formation large-scale structures \citep[see e.g.]
[for an excellent recent review on observational constraints on {\em Chameleon} modified 
gravity theories]{Lombriser_2014} -- provide several complementary ways to constrain 
deviations from standard GR. In particular, $f(R)$ gravity is already quite tightly 
constrained by the properties of small-scale structures, from the size of the Solar 
System \citep[see e.g.][]{Hu_Sawicki_2007,Lombriser_Koyama_Li_2014} to that of dwarf 
galaxies \citep[see e.g.][]{Jain_VanderPlas_2011,Jain_Vikram_Sakstein_2013,
Vikram_etal_2013}, with upper bounds on the scalar amplitude \frza\ of $8\times 
10^{-7}$  and $1\times 10^{-7}$, respectively. Despite these very tight constraints, 
$f(R)$ gravity models with larger \frza\ values still remain an interesting target 
for LSS phenomenology, as constraints on larger scales are significantly looser 
\citep[see e.g.][]{Lombriser_2014,hu16}, and since including massive neutrinos in the LSS 
modelling significantly loosens the bounds \citep{baldi14}. Furthermore, simple 
extensions of the basic $f(R)$ scenario -- e.g. by considering an effective decoupling of 
the scalar field from ordinary baryonic matter -- would evade the tightest local bounds 
and be mostly constrained by LSS formation. 

In addition to what already stated, potentially new constraints on \frz\ may be provided 
by the kinematic Sunyaev-Zel’dovich (kSZ) effect of the LSS, i.e. the Doppler-shift 
induced on the cosmic microwave background (CMB) photons by the motion of free electrons 
along the line-of-sight. In fact, the kSZ effect is expected to receive significant 
contribution by high redshift gas and is affected by motions on all physical scales 
\citep[see, e.g.,][]{roncarelli07,battaglia10,shaw12,roncarelli17}, thus being 
potentially influenced by any type of modification of the gravitational force. This 
becomes particularly interesting after the first measurement of the amplitude of  
kSZ-driven temperature fluctuations at $\ell=3000$ achieved by the South Pole Telescope 
(SPT) team. Thanks to the combination of the thermal Sunyaev Zel'dovich (tSZ) effect 
bispectrum from the SPT-SZ survey (800 deg\sqr) with the full 2540 deg\sqr\ SPT field, 
\cite{george15} obtained a $>2\sigma$ detection of the kSZ power spectrum amplitude, i.e. 
\dkszt$=(2.9\pm1.3)\,$\muk\sqr\ \citep[see also a previous measurement in a smaller area 
by][]{crawford14}. Considering that the LSS plasma at the epoch of reionization (EoR) is 
expected to provide a significant, albeit uncertain, contribution to the kSZ power 
\citep[see, e.g.,][and references therein]{iliev07,iliev14}, this value is sufficiently 
low to provide meaningful constraints on reionization models, despite the relatively 
large errors. In our previous work \citep{roncarelli17} using hydrodynamical simulations 
we predicted a kSZ power spectrum amplitude of \dkszt$=4.0$ \muk\sqr\ from the ionized 
universe in the \lcdm\ model \citep[assuming cosmological parameters from][]{planck16cp}. 
This translates into an upper limit on the contribution 
form ``patchy'' reionization of \dksztp$<1.0$ \muk\sqr\ (95 per cent C.L.), which is 
interestingly close to conservative predictions \citep[see, e.g.,][]{park13} and, 
possibly, enough to rule out some of the most extreme models \citep[see, e.g.,][]
{mesinger12,mesinger13}. In this scenario, modified gravity models are expected to 
increase the gas peculiar velocities with respect to the GR model, thus boosting the kSZ 
power, and making it a new and potentially competitive probe.\\

This paper is the second in a series of works \citep[following][Paper~I, hereafter, on 
the effect of massive neutrinos]{roncarelli17} aimed at describing the properties of the 
kSZ effect from the LSS in different cosmological models beyond the standard \lcdm. Here we 
present the first analysis of the kSZ effect signal derived from a set of cosmological 
hydrodynamical simulations that account for the effect of modified gravity with a 
state-of-the-art numerical code \citep[\mggadg, ][]
{Puchwein_Baldi_Springel_2013}. Starting from the 
physical properties of the baryons in the simulations, we derive the kSZ signal by 
constructing Doppler \bpar\ maps integrated in the past light-cone from $z=0$ down to the 
EoR, and compute the amplitude of the power spectrum ($1000 < \ell < 20000$) in the 
standard GR case and assuming different values of \frz. We also study the degeneracy 
between \zre\ (the redshift at which reionization occurs) and \frz\ and, by comparing our 
predictions with the results of \cite{george15}, we derive upper limits on \dksztp\ in 
different modified gravity scenarios.

This manuscript is organised as follows. In the next section we review the main 
definitions of modified gravity adopted for our work. In Section~\ref{s:models} we 
describe our simulations and our modelling of the kSZ effect. We discuss in 
Section~\ref{s:res} our results and draw our conclusions in Section~\ref{s:concl}.

%%%%%%%%%%%%%%%%%%%%%%%%%%%%%%%%%%%%%%%%%%%%%%%%%%%%%%%%%%%%%%%%%%%%%%%%%%%%%%%%%%%%%%%%%%
%%%%%%%%%%%%%%%%%%%%%%%%% BASIC DEFINITIONS OF MODIFIED GRAVITY %%%%%%%%%%%%%%%%%%%%%%%%%%
%%%%%%%%%%%%%%%%%%%%%%%%%%%%%%%%%%%%%%%%%%%%%%%%%%%%%%%%%%%%%%%%%%%%%%%%%%%%%%%%%%%%%%%%%%

\section{BASIC DEFINITIONS OF MODIFIED GRAVITY THEORY}
\label{s:fr}

Here we review the main definitions of the modified gravity formalism adopted in our 
work, together with the definition of \frz. For a more detailed description of this class 
of models, we remand the reader to some excellent reviews on the subject \citep[e.g.][]
{Sotiriou_Faraoni_2010,DeFelice_Tsujikawa_2010}.

As stated above, in order to pass observational tests and reproduce the observed 
expansion history the choice of the functional form of $f(R)$ must fulfil some specific 
constraints. The
most widely studied case for such constrained $f(R)$ functions is 
represented by the form \citep[][]{Hu_Sawicki_2007}:
\begin{equation}
\label{fRHS}
f(R) = -m^2 \frac{c_1 \left(\frac{R}{m^2}\right)^n}{c_2 \left(\frac{R}{m^2}\right)^n + 1},
\end{equation}
where $ m^2 \equiv H_0^2 \Omega _{\rm M}$ is a mass scale while $c_{1}$, $c_{2}$, 
{and $n$} are non-negative constant free parameters of the model. The choice of 
equation~(\ref{fRHS}) has the appealing feature of allowing to recover with arbitrary 
precision the expansion history of a $\Lambda $CDM cosmology by choosing $c_{1}/c_{2} = 
6\Omega _{\Lambda }/\Omega _{\rm M}$ under the condition $c_2 (R/m^2)^n \gg 1$, so that 
the scalar field $f_{R}$ takes the approximate form:
\begin{equation}
  f_R \approx -n \frac{c_1}{c_2^2}\left(\frac{m^2}{R}\right)^{n+1}.
\label{e:fR-R,n_relation}
\end{equation}

In the present work we will restrict our analysis to models with $n=1$, so that $c_{2}$ 
remains the only free parameter which can be also expressed in terms of the mean value 
of the scalar degree of freedom at the present epoch, $\bar{f}_{R0}$. We will then define 
our $f(R)$ cosmologies by their $\bar{f}_{R0}$ value in the following.

In $f(R)$ gravity, the dynamical gravitational potential $\Phi $ corresponding to the 
time-time metric perturbation
obeys the equation \citep{Hu_Sawicki_2007,Winther_etal_2015}:
\begin{equation}
  \nabla^2 \Phi = \frac{16 \pi G}{3} {a^2}\delta \rho - \frac{1}{6}{a^{2}}\delta R \,,
\label{e:phi_poisson_eq}
\end{equation}
which can be rewritten as:
\begin{equation}
\nabla ^{2}\Phi = \nabla ^{2}\Phi _{\rm N} -\frac{1}{2}\nabla ^{2}f_{R}
\label{e:potential}
\end{equation}
where $\Phi _{\rm N}$ is the standard Newtonian potential.
From equation~(\ref{e:potential}) it follows that the total gravitational force in 
$f(R)$ gravity is dictated by a modified potential $\Phi = \Phi_{\rm N}-\frac{1}{2}f_{R}$ 
(while the lensing potential $\Psi $ -- i.e. the space-space perturbation of the metric 
tensor -- is not affected by the modification of gravity).

The fifth-force -- proportional to $\vec{\nabla }f_{R}$ -- is suppressed for scales 
larger than the Compton wavelength of the field, which is given by
\begin{equation}
\lambda _{C} = a^{-1}\sqrt{3df_{R}/dR} = a^{-1}\sqrt{6\,|f_{R0}|\frac{\bar{R}_{0}^{2}}{R^{3}}}.
\end{equation}
This sets a maximum scale for the effects of the modification of gravity, thereby 
introducing a scale dependence in the model. Additionally, due to the nonlinear 
dependence of $\delta R$ on $f_{R}$, the theory exhibits the so-called {\em Chameleon} 
screening mechanism \citep[][]{Khoury_Weltman_2004}: the Compton wavelength 
$\lambda _{C}$ shrinks (or equivalently the scalar field mass $m_{f_{R}}\sim 
1/\lambda _{C}$ increases) in high-density environments, where the fifth-force 
propagation is then confined to arbitrarily short distances. This provides a mechanism to 
evade Solar System constraints on gravity \citep[see e.g.][]{Will_2005,
Bertotti_Iess_Tortora_2003,Lombriser_2014}.

%%%%%%%%%%%%%%%%%%%%%%%%%%%%%%%%%%%%%%%%%%%%%%%%%%%%%%%%%%%%%%%%%%%%%%%%%%%%%%%%%%%%%%%%%%
%%%%%%%%%%%%%%%%%%% MODELLING THE EFFECT OF f(R) ON THE KSZ SIGNAL %%%%%%%%%%%%%%%%%%%%%%%
%%%%%%%%%%%%%%%%%%%%%%%%%%%%%%%%%%%%%%%%%%%%%%%%%%%%%%%%%%%%%%%%%%%%%%%%%%%%%%%%%%%%%%%%%%

\section{MODELLING THE EFFECT OF $f(R)$ ON THE kSZ SIGNAL}
\label{s:models}

\subsection{The hydrodynamical simulations}
\label{ss:hydro}

As mentioned above, for our simulations we have used the \mggadg\ code \citep[][]
{Puchwein_Baldi_Springel_2013}, a modified version of \gadgetiii\ \citep[][]{
springel05} that includes the effects of the modified potential and its associated 
{\em Chameleon} screening mechanism. \mggadg\ solves 
equation~(\ref{e:fR_field_eq}) for a generic density field produced by a set of discrete 
particles by means of a  Newton-Gauss-Seidl iterative scheme, and computes the total 
force on each particle via equation~(\ref{e:phi_poisson_eq}) by including the curvature 
perturbation $\delta R$ (derived according to equation~\ref{e:fR_field_eq}) in the 
gravitational source term. We refer the interested reader to the \mggadg\ code 
paper \citep[][]{Puchwein_Baldi_Springel_2013} for an extended presentation of the 
numerical implementation.

In the present work we will consider 4 different hydrodynamical simulations, whose 
characteristics are summarized in Table~\ref{t:sim}. The parameters have been chosen 
to mimic exactly the configuration of our first simulation set of Paper~I (dubbed N0, 
N15, N30 and N60) on the effect of massive neutrinos, to allow for a direct comparison.
Besides the GR simulation, that corresponds to the standard \lcdm\ scenario (\frz=0), we 
run three other simulations, dubbed FR-6, FR-5 and FR-4, with \frztnz, respectively. 
While the largest of these values is already ruled out (see Section~\ref{s:intro}) by 
other probes (even though it was never tested before using the kSZ effect alone, as we 
will do in our analysis) the other models considered in the present work are still 
marginally consistent with the most recent observational constraints and are widely 
employed by the community to test $f(R)$ gravity phenomenology. \\

Our simulations evolve a periodic distribution of $512^{3}$ CDM particles and an equal 
number of baryonic particles, in a cosmological box of $240\, h^{-1}$ Mpc per side, from 
a starting redshift of $z_{i}=99$. Initial conditions have been generated by displacing 
particles from a homogeneous cartesian grid according to the Zel'dovich approximation 
\citep[][]{Zeldovich_1970} in order to obtain a random realisation of the linear matter 
power spectrum computed with the public Boltzmann code \textsc{camb} \citep[][]{CAMB} for 
the set of cosmological parameters summarised in the caption of Table~\ref{t:sim}. As the 
effects of Modified Gravity are fully screened in the high-redshift Universe by the high 
background density, no modification is needed in such procedure to generate initial 
conditions for the $f(R)$ gravity simulations. Thus, all our simulations use exactly the 
same starting configuration at $z_{i}=99$, and evolve according to different gravity 
theories from then on. CDM particles are treated as a collisionless fluid, conversely the 
thermo- and hydrodynamics of baryons is modelled through the Smoothed Particle 
Hydrodynamics (SPH) scheme included in \gadgetiii. Besides adiabatic hydrodynamical 
forces, the treatment of SPH particles employs the simplified `quick-Lyman-$\alpha$' 
approach \citep[][]{Viel_etal_2005} that amounts to convert into a collisionless star 
particle every gas particle with density contrast  above $1000$ and temperature below 
$10^{5}$ K. Such treatment is much less numerically demanding compared to more 
sophisticated implementations of star formation and feedback mechanisms, but is largely 
sufficient for the purposes of the present work.

\begin{table}
\begin{center}
\caption{
  Parameter values of our set of simulations. First column: simulation name. Second 
  column: comoving box size, in \hmone\ Mpc. Third column: total number of particles 
  (DM + baryons). Fourth and fifth column: masses of the DM and baryonic particles, 
  respectively, in units of $10^{9}$\hmone\msun. Sixth column: value of \frz. All 
  simulations assume a flat cosmology with \omegal$=0.6866$, 
  \omegam$=$\omegacdm$+$\omegab$=0.3134$, \omegab$=0.049$, \hzero$=67$ km/s/Mpc 
  ($h=0.67$), $n_s=0.96$ and $A_s=2.13\times10^{-9}$ (corresponding to \sigmae=0.834 at 
  $z$=0 for the GR simulation). The physical scheme assumed for the baryonic component 
  includes radiative cooling, UV background and star formation following the 
  `quick-Lyman-$\alpha$' method (see text for details). These parameters 
  are defined so that the GR simulation matches the N0 simulation of Paper~I.
  }
\begin{tabular}{rcccccc}
\hline
\hline
Simulation && $L_{\rm box}$ & $N_p$ & $m_{p, \rm cdm}$ & $m_{p,\rm b}$ & \frz  \\
           && (\hmone\ Mpc) &       & \multicolumn{2}{c}{($10^{9}$\hmone\msun)} & \\
\hline
  GR && 240 & 2$\times$  512\cub & 7.56  & 1.40 &       $0$  \\
FR-6 && 240 & 2$\times$  512\cub & 7.56  & 1.40 & $-10^{-6}$ \\
FR-5 && 240 & 2$\times$  512\cub & 7.56  & 1.40 & $-10^{-5}$ \\
FR-4 && 240 & 2$\times$  512\cub & 7.56  & 1.40 & $-10^{-4}$ \\
\hline
\hline
\label{t:sim}
\end{tabular}
\end{center}
\end{table}

Since it does not depend directly on gas temperature (see details in 
Section~\ref{ss:ksz}), the kSZ effect of the ionised universe is weakly dependent by the 
details of the feedback mechanisms affecting the baryonic component \citep[see, e.g.,]
[and discussions therein]{roncarelli07,trac11,shaw12}, unlike, for instance, the X-ray 
emission or the tSZ effect. When modelling the kSZ effect with hydrodynamical simulations 
one has to face two main challenges: {\it (i)} accounting for the amount of baryons that 
does not reside in the diffuse ionised state (mainly stars) and {\it (ii)} sampling a 
cosmological volume big enough to sample the large-scale velocity modes. In this work we 
use the same approach adopted in Paper~I: we refer the interested reader to its Section~4 
for the details of the implementation and for the relative tests with specific 
hydrodynamical simulations used to quantify the systematics. We summarise here the main 
points.

\begin{figure}
\includegraphics[width=0.5\textwidth]{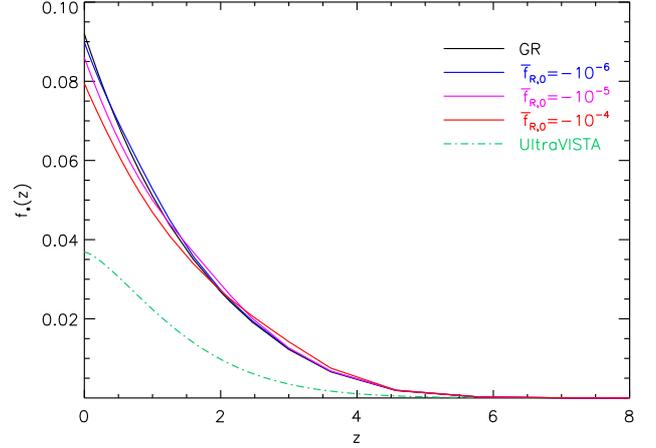}
\caption{
  Global star mass fraction ($\rho_\star / \rho_{\rm b}$) in the whole box volume as a 
  function of redshift for our set of four simulations (solid lines), with \frzt\ 
  shown in black, blue, magenta and red, respectively. The green dot--dashed line is 
  derived from the cosmic stellar mass density estimated by \protect\cite{ilbert13}.
  }
\label{f:fstar}
\end{figure}

\begin{enumerate}
\item We compute the star mass fraction as a function of redshift in our simulation 
volumes, $f_{\star, \rm sim}(z)$, shown in Fig.~\ref{f:fstar}. Since our cooling scheme 
is known to overpredict this quantity, we take as a reference the cosmological star mass 
fraction, $f_{\star, \rm obs}(z)$, derived by \cite{ilbert13} with UltraVISTA data and 
correct the density of our SPH particles in the following way:
\begin{equation}
\rho_{\rm g} (z) = 
\rho_{\rm g,sim}(z)\left[\frac{1-f_{\star,\rm obs}(z)}{1-f_{\star,\rm sim}(z)}\right]\, .
\label{e:rhoc}
\end{equation}
As it can be seen from the plot in Fig.~\ref{f:fstar}, this corresponds to correcting the 
density by $\sim 1$ per cent at $z=3$ and $\sim 5$ per cent at $z=0$. We have shown in 
Paper~I that this approach allows to correct the kSZ 
signal mimicking the desired star formation history without introducing significant 
systematics (less than 5 per cent at $\ell < 5000$).
\item Given the limited volume of our simulations, we need to correct for the missing 
velocity power on scales larger than our box size of 240\hmone\ Mpc. By adopting an 
approach similar (albeit slightly more conservative) to \cite{iliev07}, we have shown in 
Paper~I (Section 4) that we need to increase the amplitude of the kSZ power computed from 
our simulations by 25 per cent\footnote{In Paper~I we quoted the value of 20 per cent, 
referring to the missing power with respect to the total one: this actually corresponds 
to increase by 25 per cent the simulations results.} in order to account for the missing 
velocity power. Since this paper has its main focus in studying the impact of modified 
gravity on the kSZ effect, we stress that the results presented here on the relative 
difference between our 4 simulations do not depend on this correction. However, this 
point clearly becomes crucial when comparing our results with SPT data, as we do in 
Section~\ref{s:res}.
\end{enumerate}

%%%%%%%%%%%%%%%%%%%%%%%%%%%%%%%%%%%%%%%%%%%%%%%%%%%%%%%%%%%%%%%%%%%%%%%%%%%%%%%%%%%%%%%%%%
\subsection{Light-cone construction} %%%%%%%%%%%%%%%%%%%%%%%%%%%%%%%%%%%%%%%%%%%%%%%%%%%%%
\label{ss:lcone}

The study of the kSZ properties requires a full light-cone reconstruction, accounting for 
the integrated signal up to the EoR (see Section~\ref{ss:ksz}). Here we review our method 
described also in Paper~I and already adopted in our previous works \citep{roncarelli06a,
roncarelli07,roncarelli10a,roncarelli12,roncarelli15}.

We stack the simulation outputs along the line of sight up to $z=15$, enough to account 
for the most conservative upper limits in \zre. With our cosmology, this corresponds to 
a comoving distance of 7022\hmone\ Mpc. The positions of the SPH particles inside each 
volume are randomized to avoid the repetition of the same structure inside the field of 
view (see details in Paper~I). The randomization process is implemented using the same 
random seed for all simulations. Since they also share the phases of the initial 
conditions, this ensures that the volumes enclosed by the light-cones reproduce an 
identical realization of the same structures, albeit with different \frz, thus 
eliminating the effect of cosmic variance in the relative differences. By varying the 
initial random seed, we generate 50 light-cone realizations for each simulation: this 
allows to enhance the statistical robustness of our final results.

%%%%%%%%%%%%%%%%%%%%%%%%%%%%%%%%%%%%%%%%%%%%%%%%%%%%%%%%%%%%%%%%%%%%%%%%%%%%%%%%%%%%%%%%%%
\subsection{The kSZ effect model} %%%%%%%%%%%%%%%%%%%%%%%%%%%%%%%%%%%%%%%%%%%%%%%%%%%%%%%%
\label{ss:ksz}

The kSZ effect \citep{sunyaev70,ostriker86,vishniac87} is the Doppler shift in the CMB 
spectrum induced by the peculiar velocity of the free electrons of the LSS. Unlike for 
its thermal counterpart, the shift, $\Delta T$, in the observed CMB temperature is the 
same at all frequencies: for a given direction, identified by the unit vector 
$\pmb{\hat\gamma}$, one has
\begin{equation}
\Delta T(\pmb{\hat\gamma}) = -b(\pmb{\hat\gamma}) \, T_{\rm CMB} \, ,
\label{e:dt}
\end{equation}
where $T_{\rm CMB}$ is the CMB temperature and $b(\pmb{\hat\gamma})$ is the Doppler 
\bpar, defined as
\begin{equation}
b(\pmb{\hat\gamma}) \equiv 
\frac{\sigma_{\rm T}}{c} \int_0^{z_{\rm re}} \! \frac{{\rm d}\chi}{{\rm d}z} \frac{{\rm d}z}{1+z} \, 
e^{-\tau(z)} \, n_{\rm e} \ \pmb{v}_{\rm e} \cdot \pmb{\hat\gamma} \, .
\label{e:bdef}
\end{equation}
In the previous equation $c$ is the vacuum light speed and $\sigma_{\rm T}$ the Thomson 
cross-section. The intergalactic medium (IGM) physical properties enter in the integral 
along the comoving coordinate $\chi$, with $n_{\rm e}$ and $\pmb{v}_{\rm e}$ being the 
number density and the proper velocity of the electrons, respectively. Finally, the 
Thomson optical depth, $\tau$, is given by:
\begin{equation}
\tau(z) \equiv \sigma_{\rm T}\int_0^{z}\!\frac{{\rm d}\chi}{{\rm d}z'}\frac{{\rm d}z'}{1+z'}\,n_{\rm e}(z') \, .
\label{e:tau}
\end{equation}

As already done in Paper~I, in order to compute equation~(\ref{e:bdef}) from our 
\gadgetiii\ simulations, we apply the following equation:
\begin{equation}
b(\pmb{\hat\gamma}) = \frac{\sigma_{\rm T}X x_{\rm e}}{c \, m_{\rm p}} \int_0^{z_{\rm re}} \! 
                 \frac{{\rm d}\chi}{{\rm d}z} \frac{{\rm d}z}{1+z} \, e^{-\tau(z)} \, \rho_{\rm g} \ 
                 \pmb{v}_{\rm g} \cdot \pmb{\hat\gamma} \, ,
\label{e:bapp}
\end{equation}
where $X=0.76$ is the cosmological hydrogen mass fraction, $x_{\rm e} \simeq 1.16$ is the 
electron-to-proton ratio and $m_{\rm p}$ the proton mass. As said in 
Section~\ref{ss:hydro}, to account for the fraction of baryons that are not ionised, we 
correct the value of the gas mass density, $\rho_{\rm g}$, with equation~(\ref{e:rhoc}).

%%%%%%%%%%%%%%%%%%%%%%%%%%%%%%%%%%%%%%%%%%%%%%%%%%%%%%%%%%%%%%%%%%%%%%%%%%%%%%%%%%%%%%%%%%
\subsection{The mapping procedure} %%%%%%%%%%%%%%%%%%%%%%%%%%%%%%%%%%%%%%%%%%%%%%%%%%%%%%%
\label{ss:maps}

The kSZ physical model described in Section~\ref{ss:ksz} and the light-cone geometry 
defined in Section~\ref{ss:lcone} are employed to create a set of \bpar\ maps, that are 
then used to compute our results. Here we summarize the main points of our method.

For a given SPH particle, the physical variables provided by \mggadg\ that are relevant 
for the kSZ effect are its mass, $m_i$, its peculiar velocity, $\bf v_i$, and its 
smoothing length, $h_i$. In addition, its 3D position in the light-cone determines its 
sky coordinate $\pmb{\hat\gamma}_i$ that is used both to place it in the map and to 
determine its radial velocity component $v_{r,i}\equiv \bf v_i \cdot \pmb{\hat\gamma}_i$.  
After selecting all the particles whose `SPH-sphere', i.e. the sphere with radius $h_i$, 
intersects the light-cone, their integrated \bpar\ is computed with the following 
formula:
\begin{equation}
B_i = 
\frac{X\,\sigma_{\rm T}\,x_{\rm e}}{m_{\rm p}\,c\,d_{A,i}^2} \, 
e^{-\tau(z_i)}\,m_i\,v_{r,i} \, ,
\label{e:bsph}
\end{equation}
being $d_{A,i}$ the angular diameter distance from the observer, and $z_i$ the 
corresponding cosmological redshift, used to compute $\tau(z_i)$ with 
equation~(\ref{e:tau}). Finally, equation~(\ref{e:bapp}) is converted into a 
discrete sum over all the values of $B_i$, that are distributed in the map pixels 
exploiting the SPH mathematical formalism to ensure both the accuracy and the 
computational performance of the calculation\footnote{The interested reader may refer to 
\cite{ursino10} for the details of the numerical implementation of the SPH mapping 
procedure.}.

For our purposes, we compute maps of 1.8$^\circ$ per side: at $z=15$ the corresponding 
transverse comoving distance is 221\hmone\ Mpc, so that our box size of 240\hmone\ Mpc 
fully encloses the field of view. Each map is 1024 pixels per side, corresponding to an 
angular resolution of 6.33 arcsec. The mapping procedure is repeated for the 50 
light-cones and for the full set of simulations. Moreover, we compute separately the 
contribution to each \bpar\ map into 20 logarithmically equi-spaced redshift bins, 
by varying the limits in the integral of equation~(\ref{e:bapp}). On the whole this makes 
a total of 1000 \bpar\ maps for each simulation.

%%%%%%%%%%%%%%%%%%%%%%%%%%%%%%%%%%%%%%%%%%%%%%%%%%%%%%%%%%%%%%%%%%%%%%%%%%%%%%%%%%%%%%%%%%
%%%%%%%%%%%%%%%%%%%%%%%%%%%%%%%%%%%%%%%%% RESULTS %%%%%%%%%%%%%%%%%%%%%%%%%%%%%%%%%%%%%%%%
%%%%%%%%%%%%%%%%%%%%%%%%%%%%%%%%%%%%%%%%%%%%%%%%%%%%%%%%%%%%%%%%%%%%%%%%%%%%%%%%%%%%%%%%%%

\section{RESULTS}
\label{s:res}

%%%%%%%%%%%%%%%%%%%%%%%%%%%%%%%%%%%%%%%%%%%%%%%%%%%%%%%%%%%%%%%%%%%%%%%%%%%%%%%%%%%%%%%%%%
\subsection{Global properties of the Doppler \bpar} %%%%%%%%%%%%%%%%%%%%%%%%%%%%%%%%%%%%%%
\label{ss:glob}

\begin{figure*}
\vspace{-5cm}
\includegraphics{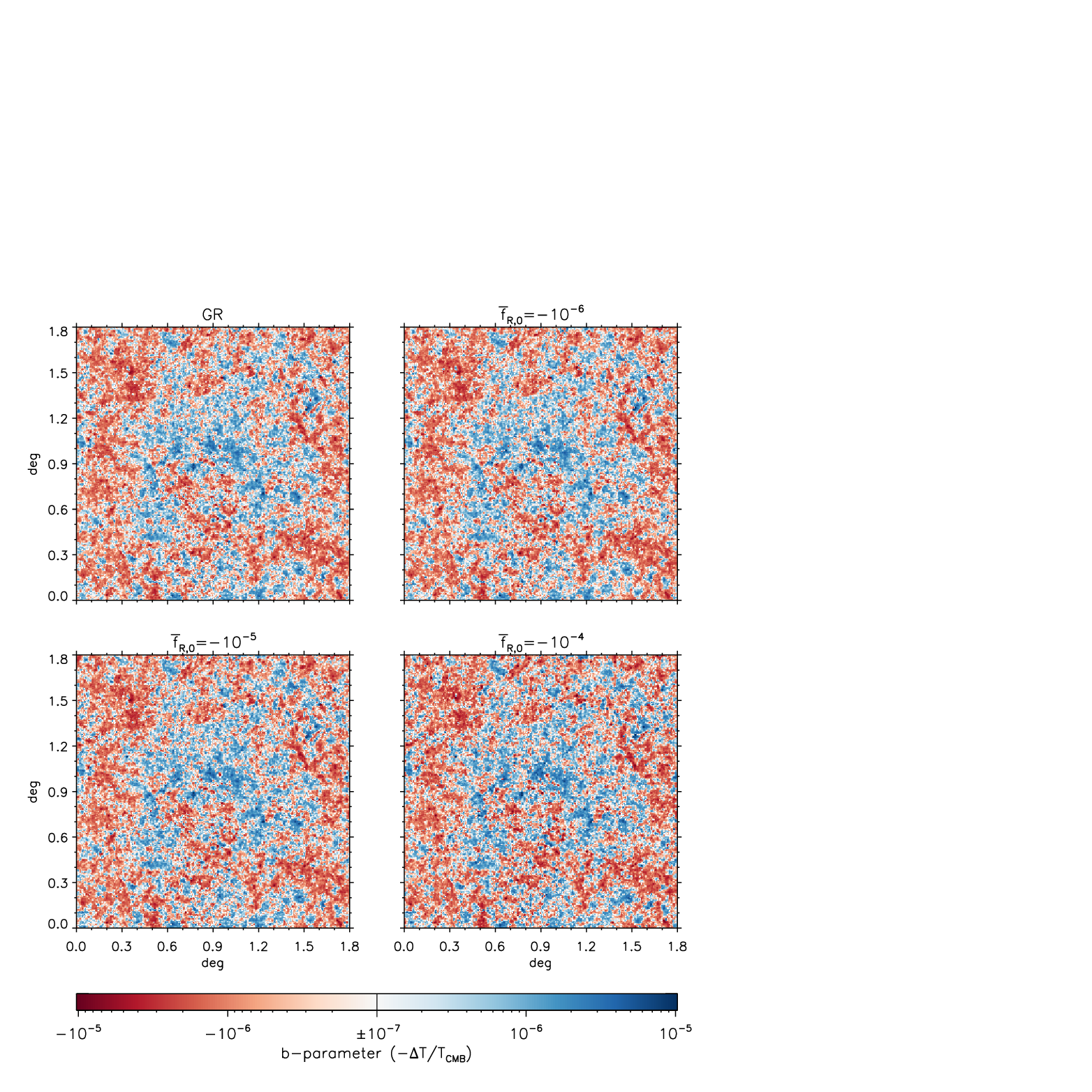}
\caption{
  Maps of the \bpar\ as a function of \frz. Each map is 1.8$^\circ$ per side with a 
  resolution of 6.33 arcsec ($1024^2$ pixels) and represents the signal integrated 
  from $z=0$ to \zre$=8.8$ for the same light-cone assuming general relativity (\frz$=0$, 
  top-left) and \frztnz\ (top-right, bottom-left and bottom-right, respectively). 
  The color scale indicates in red the sky regions where, on average, the gas is 
  approaching the observer ($b<0$ and increase of observed CMB temperature), and in blue 
  where the gas is receding ($b>0$ and decrease of observed CMB temperature).
  }
\label{f:maps}
\end{figure*}

We show in Fig.~\ref{f:maps} the \bpar\ maps for our 4 simulations with \frzt. The maps 
have been obtained from the same light-cone and considering \zre$=8.8$, i.e. the nominal 
value as measured by \cite{planck16cp}. Coherently with our analysis of Paper~I, the 
typical values are of the order of $|b| \approx 10^{-6}$, enough to induce CMB temperature 
increments or decrements of several \muk\ (see equation~\ref{e:dt}). The map peaks, 
associated to galaxy clusters, can reach values an order of magnitude larger. The larger 
modes of the size of $\sim$10 arcmin are connected to coherent motions in the LSS.

\begin{figure}
\includegraphics[width=0.5\textwidth, trim=0.5cm 0 -0.5cm 0]{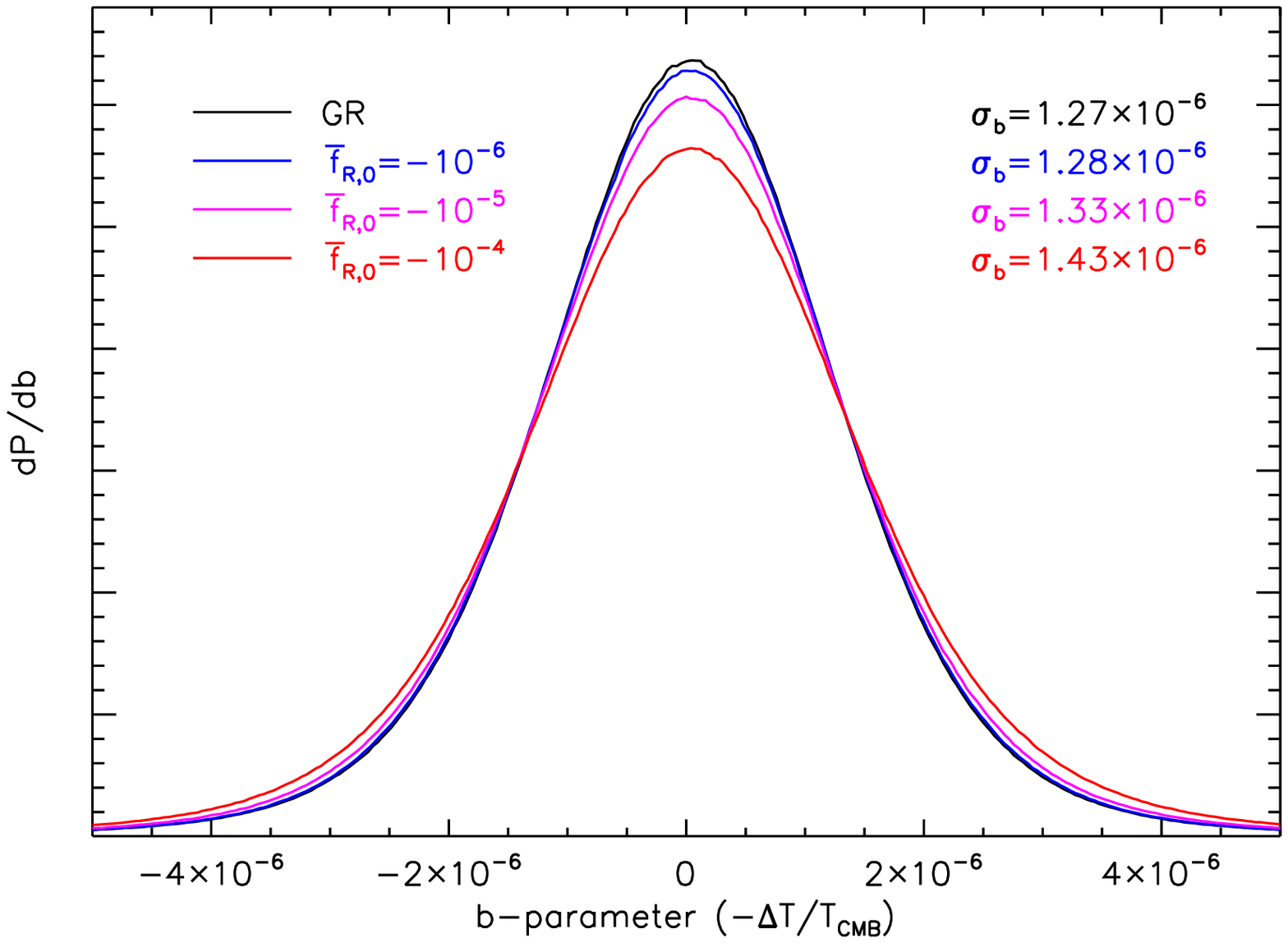}
\caption{
  Probability distribution function of the \bpar\ for different values of \frz. Each 
  curve is computed for the whole set of 50 light cones (i.e. 50 sets of maps like the 
  ones shown in Fig.~\ref{f:maps}) with signal integrated up to \zre$=8.8$ and pixels of 
  6.33 arcsec per side. The color coding is the same as in Fig.~\ref{f:fstar}. On the 
  top right corner we show the standard deviation of the best-fitting Gaussian 
  distribution of each curve.
  }
\label{f:bdist}
\end{figure}

The effect of the different values of \frz\ is not evident by eye. In order to highlight 
these differences, we show in Fig.~\ref{f:bdist} the distribution of pixel values 
computed for the 4 models from the full 50 light-cones (i.e. a total of $5.2 \times 
10^7$ pixels each). It is clear that the effect of $f(R)$ models is to increase the 
chance of larger \bpar\ values: qualitatively this is expected, since the enhancement of 
the gravitational force with respect to the standard GR model has a direct impact on the 
velocity field. The effect increases for larger absolute values of \frz.

We fit all the 4 distributions with a Gaussian curve, with fixed normalization and with 
mean and dispersion, $\sigma_b$, as free parameters. The results for the values of 
$\sigma_b$, shown in the top right corner of Fig.~\ref{f:bdist}, range from $1.27 \times 
10^{-6}$ for the GR model to $1.43 \times 10^{-6}$ for \frz$=-10^{-4}$. We have verified 
empirically that the scaling of $\sigma_b$ with \frz\ well follows a power-law form of 
this type:

\begin{equation}
\sigma_{b,\overline{f}_{\rm R,0}} = \sigma_{b,{\rm GR}} \, \left(1 +\sqrt{ \left| \overline{f}_{\rm R,0} \right|}\right)^{12.5} \, ,
\label{e:sbsc}
\end{equation}
where $\sigma_{b,{\rm GR}}$ is the value for the GR model. As we will show in 
Section~\ref{ss:pows}, the dependence with $\left(1 +\sqrt{ \left| \overline{f}_{\rm R,0} 
\right|}\right)$ does not describe only the scaling of 
$\sigma_{b}$, but works well also to trace the variations of the kSZ power spectrum 
amplitude. Since this is the attempt to define the scaling of an observational quantity 
in $f(R)$ models, a possible interesting test would be to verify if a formula of this 
type works also with other important physical or observational quantities, such as the 
halo mass function or the tSZ effect power spectrum. Since this goes beyond the scope of 
the present paper, we leave this for future works.

%%%%%%%%%%%%%%%%%%%%%%%%%%%%%%%%%%%%%%%%%%%%%%%%%%%%%%%%%%%%%%%%%%%%%%%%%%%%%%%%%%%%%%%%%%
\subsection{The kSZ effect power spectrum} %%%%%%%%%%%%%%%%%%%%%%%%%%%%%%%%%%%%%%%%%%%%%%%
\label{ss:pows}

In order to compare our results with the SPT measurement by \cite{george15} we need to 
calculate the amplitude of the angular power spectrum of temperature fluctuations induced 
by the kSZ effect as a function of the multipole $\ell$. We do so by converting our 
\bpar\ into $\Delta T$ maps with equation~(\ref{e:dt}) and by applying a Fast Fourier 
Transform in the flat-sky approximation \citep[see the detailed explanation in][]
{roncarelli07}. Following \cite{crawford14}, we express the results in terms of 
\begin{equation}
\mathcal{D}_{\ell} \equiv \frac{\ell(\ell+1)\,C_{\ell}}{2\upi} \, ,
\end{equation}
where $C_{\ell}$ is the definition of power spectrum of temperature fluctuations normally 
adopted in CMB analyses.

\begin{figure}
\includegraphics[width=0.5\textwidth]{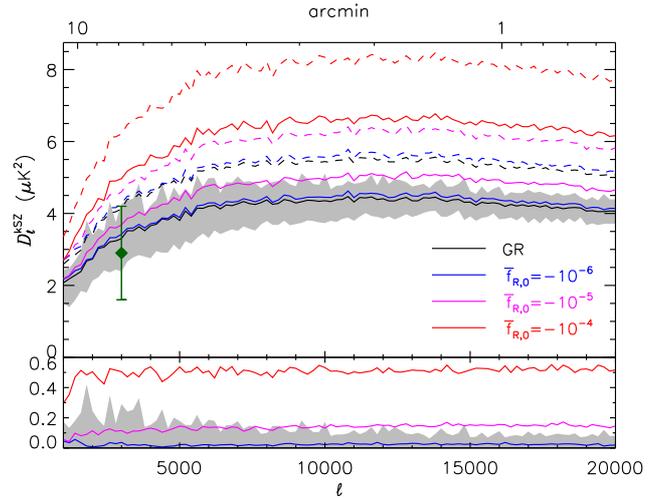}
\caption{
  Top panel: angular power spectrum of temperature fluctuations associated with the kSZ 
  effect (\dksz) as a function of the multipole $\ell$ for different values of \frz. 
  The curves are computed by averaging over the full set of 50 light-cones (a total of 
  162 deg\sqr) and consider the signal integrated up to \zre$=8.8$. The color coding is 
  the same as in Fig.~\ref{f:fstar}. The grey-shaded area encloses the 
  central 34 light-cones (68 per cent of the total) for the GR model (\frz$=0$) only. 
  Solid lines indicate values derived directly from the maps, dashed lines account for 
  the 20 per cent correction due to the limited box size. The 
  green diamond with error bars shows the measurement of \dkszt$=(2.9\pm1.3)$ \muk\sqr\ 
  by \protect\cite{george15}. Bottom panel: relative differences with respect to the GR 
  model.
  }
\label{f:pows}
\end{figure}

The results of the kSZ power spectrum amplitude as a function of $\ell$ for our 4 models 
and assuming \zre$=8.8$ are shown in Fig.~\ref{f:pows} (solid lines). Overall the shape 
of the power spectrum is the same for all models, with an increase from $\ell=1000$ to 
6000 followed by an almost flat ``plateau'': this is consistent with our previous results 
in Paper~I and in \cite{roncarelli07}. As expected, the effect of modified gravity is to 
boost the kSZ power: this enhancement is 2--3 per cent for the FR-6 model, 10--15 per 
cent for the FR-5, up to about 50 per cent for the FR-4. Interestingly, the effect of 
$f(R)$ appears  to be scale-independent, i.e. the relative differences do not depend on 
$\ell$. On the other hand, the effect of massive neutrinos discussed in Paper~I showed a 
clear dependence on the angular scale, decreasing in intensity for larger $\ell$, thus 
providing a potential method to remove the degeneracies between the two effects. 
In fact, it is now well known that $f(R)$ gravity and massive neutrinos are strongly 
degenerate with each other, so that specific combinations of their characteristic 
parameters (namely \frz\ and $\Sigma m_\nu$) may be hardly distinguishable 
from the standard $\Lambda $CDM model for a wide range of basic cosmological observables 
\citep[see e.g.][]{He_2013,baldi14,Motohashi_etal_2013,Wright_Winther_Koyama_2017} and 
more sophisticated statistics are needed to disentangle their effects \citep[see e.g. the 
recent outcomes of][]{Peel_etal_2018}.

Apart from relative differences, a comparison with 
observational results requires to apply the correction due to the limited simulation 
volume detailed in Section~\ref{ss:hydro}, that we estimate in an increase of 25 per 
cent. The corresponding curves with this corrections are shown with dashed lines in 
Fig.~\ref{f:pows}. Most importantly, we predict a value of \dkszt$=4.1\,$\muk\sqr\ for 
the \lcdm-GR model, consistent with our previous result of $=4.0\,$\muk\sqr\ in Paper~I 
(N0 run, volume corrected), enough to account for all of the signal of 
\dkszt$=(2.9\pm1.3)\,$\muk\sqr\ measured by \cite{george15}, and close to its $+1\sigma$ 
limit, even without including the contribution from patchy reionization. We discuss the 
implications of our result on these type of models in Section~\ref{ss:impl}. As 
expected, our three modified gravity models predict larger values, with 
\dkszt$=(4.3,4.6,6.2)\,$\muk\sqr\ for \frztnz, respectively. Only the \frz$=-10^{-4}$ 
model is ruled out at a significant level ($2.5\sigma$), indicating that differences 
associated with realistic $f(R)$ model would be difficult to measure with current 
CMB instruments.

\begin{figure}
\includegraphics[width=0.5\textwidth]{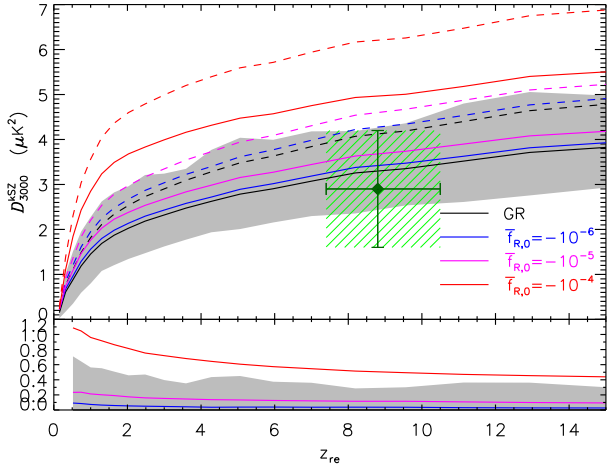}
\caption{
  Top panel: amplitude of the kSZ power spectrum at $\ell=3000$ (\dkszt) as a function of 
  \zre\ for different values of \frz. The color coding is the same as in 
  Fig.~\ref{f:fstar}. The grey-shaded area encloses the central 34 light-cones (68 per 
  cent of the total) for the GR model (\frz$=0$) only. Solid lines indicate values 
  derived directly from the maps, dashed lines account for the 25 per cent correction due 
  to the limited box size. The green diamond with error bars (1$\sigma$) and 
  green-shaded area shows the results of \zre$=8.8^{+1.7}_{-1.4}$ and 
  \dkszt$=(2.9\pm1.3)$ \muk\sqr\ by \protect\cite{planck16cp} and 
  \protect\cite{george15}, respectively. Bottom panel: relative differences with respect 
  to the GR model.
  }
\label{f:zre}
\end{figure}

The predictions on the kSZ effect observables depend on the cosmological assumptions 
\citep{shaw12}. Therefore our results on \frz\ are degenerate with respect to the 
standard cosmological parameters and with the value of the neutrino mass fraction, \fnu, 
as studied in Paper~I. With our set of data it is easy to show the dependence of our 
results on the value of \zre\ that, given the current measurements errors, constitutes 
one of the main sources of uncertainty.

Fig.~\ref{f:zre} shows the values of \dkszt\ as a function of \zre\ for our 4 simulations 
(solid lines) obtained by varying iteratively the upper limit in the integral of 
equation~(\ref{e:bapp}). We also show with dashed lines the volume corrected values that 
can be directly compared with the SPT measurements, with uncertainties, by 
\cite{george15} and the estimate of \zre\ by \cite{planck16cp} (green point with shaded 
region). These curves show how the kSZ effect receives significant contribution from both 
low and high redshift gas, as already discussed in Paper~I \citep[see also][]
{roncarelli07,shaw12}. Most notably, we observe that the differences induced by modified 
gravity are particularly relevant at later epochs, where non-linear effects play an 
important role. As an example, in the $0<z<1$ range the value of \dkszt\ for the FR-4 
model is double with respect to the GR one: 1.8 and 3.6 \muk\sqr\ (volume corrected), 
respectively. On the contrary, differences in the contribution at earlier epochs are 
less pronounced.

\begin{table}
\begin{center}
\caption{
  Amplitude of the kSZ effect power spectrum, \dksz, and its dependence with \zre\ and 
  \frz. First column: multipole $\ell$. Second column: value of \dksz, in \muk\sqr, 
  obtained assuming GR (\frz$=0$), computed averaging over the 50 light-cones (3.24 
  deg\sqr\ each). Third column: \dksz\ after applying a correction that accounts for the
  25 per cent correction (i.e. 20 per cent of missing signal) due to the limited box 
  size. The last two columns show the dependence of \dksz\ on the redshift of 
  reionization and \frz, in terms of best-fit values of the exponents $\alpha$ and 
  $\gamma$ as in equation~(\ref{e:fit}).
  }
\begin{tabular}{ccccccc}
\hline
\hline
$\ell$ &&  \multicolumn{2}{c}{\dksz\ $\left(\umu{\rm K}^2\right)$} && 
\multicolumn{2}{c}{$\left(\frac{z_{\rm re}}{8.8}\right)^\alpha \, \left(1+\sqrt{\left|\overline{f}_{\rm R,0}\right|}\right)^\gamma$}  \\
\cline{3-4} \cline{6-7}
       &&  Uncorr. & Vol. corr. &&  $\alpha$ & $\gamma$ \\
\hline
 2000 && 2.73 & 3.4 && 0.30 & 41 \\
 3000 && 3.30 & 4.1 && 0.28 & 41 \\
 4000 && 3.66 & 4.6 && 0.24 & 43 \\
 5000 && 3.80 & 4.8 && 0.22 & 43 \\
 6000 && 4.20 & 5.2 && 0.21 & 41 \\
 8000 && 4.37 & 5.5 && 0.17 & 41 \\
10000 && 4.35 & 5.4 && 0.19 & 43 \\
15000 && 4.31 & 5.4 && 0.15 & 42 \\
20000 && 4.06 & 5.1 && 0.13 & 42 \\
\hline
\hline
\label{t:fit}
\end{tabular}
\end{center}
\end{table}

Following our analysis in Paper~I, we provide a more direct estimate of the dependence 
with respect to \zre\ and \frz\ by defining an analytical expression. We observe that 
our results can be fit with the following formula:
\begin{equation}
\mathcal{D}^{\rm kSZ}_{\ell} = 
\mathcal{D}^{\rm kSZ}_{{\rm GR},\ell} \, 
\left(\frac{z_{\rm re}}{8.8}\right)^\alpha \, 
\left(1+\sqrt{\left|\overline{f}_{\rm R,0}\right|}\right)^\gamma \, ,
\label{e:fit}
\end{equation}
where $\mathcal{D}^{\rm kSZ}_{{\rm GR},\ell}$ is the result for our GR model. The 
best-fit values of $\alpha$ and $\gamma$ for the different multipoles are shown in 
Table~\ref{t:fit}, together with the corresponding values of \dksz, with and without 
volume correction. These results confirm our previous findings of Paper~I (see Table 2), 
with the values of \dksz\ consistent in the range of a few percent, and with almost 
identical values of $\alpha$. This corroborates the fact that the dependence on \zre\ 
decreases substantially at smaller angular scales, going from $\alpha=0.30$ at 
$\ell=2000$, down to $0.13$ at $\ell=20000$. The results on the scaling with \frz\ 
confirm in a more quantitative way to be scale-independent, with scaling in the range 
$\left(1+\sqrt{\left|\overline{f}_{\rm R,0}\right|}\right)^{41-43}$ with no 
significant trend with $\ell$. Since the scaling with \sigmae\ is expected to vary, 
albeit mildly, with angular scale \citep[see, e.g.,][]{shaw12}, this suggests that the 
combination of \dksz\ measurements at different $\ell$ may break the \sigmae--\frz\ 
degeneracy.

Finally, by combining the results presented here with the ones of Paper~I we can provide 
a comprehensive formula for the scaling of the amplitude of the kSZ power spectrum 
for a combination of $f(R)$ modified gravity and massive neutrinos\footnote{The formula 
that appears in equation~(11) of Paper~I mistakenly reports \zre$^\alpha$ instead of the 
correct $\left(\frac{z_{\rm re}}{8.8}\right)^\alpha$.}:

\begin{equation}
\mathcal{D}^{\rm kSZ}_{\ell} = 
\mathcal{D}^{\rm kSZ}_{0,\ell} \, 
\left(\frac{z_{\rm re}}{8.8}\right)^\alpha \,
(1-f_\nu)^\beta \,
\left(1+\sqrt{\left|\overline{f}_{\rm R,0}\right|}\right)^\gamma \, .
\label{e:fittot}
\end{equation}
This expression accounts for the dependence on \zre, the neutrino mass fraction \fnu\ 
(see the values of $\beta$ in Table~2 of Paper~I) and modified gravity through \frz, 
and is valid from $\ell=2000$ to $20000$, under the assumption that the two effects are 
decoupled. Although the last statement seems to be supported by the tight degeneracy 
between the respective observational effects \citep[see again][]{baldi14,Peel_etal_2018} 
it was certainly not yet demonstrated for kSZ observations, which would require running a 
dedicated suite of combined hydrodynamical simulations. This goes beyond the scope of the 
present paper, and we leave such analysis for future work.

%%%%%%%%%%%%%%%%%%%%%%%%%%%%%%%%%%%%%%%%%%%%%%%%%%%%%%%%%%%%%%%%%%%%%%%%%%%%%%%%%%%%%%%%%%
\subsection{Comparison with the literature} %%%%%%%%%%%%%%%%%%%%%%%%%%%%%%%%%%%%%%%%%%%%%%
\label{ss:comp}

The effect of modified gravity on the kSZ power spectrum is a pioneer topic, that has not 
been much explored in the literature. The only work so far is by \cite{bianchini16}, who 
provide a set of analytical estimates of the post-ionization kSZ power spectrum shape 
with a family of modified gravity models similar to the ones adopted here, i.e. following 
the \cite{Hu_Sawicki_2007} formalism with $n=1$. A comparison of our works is therefore 
very interesting, in particular to test the possibile differences associated to their 
analytical approach, using power spectra obtained with \textsc{mg-camb} \citep{zhao09} 
and \textsc{mg-halofit} \citep{zhao14}, with respect to our numerical method with \mggadg.

In agreement with our results, \cite{bianchini16} observe an increase of the kSZ that 
does not depend on the angular scale. This confirms that the scale dependence induced in 
the density and velocity power spectrum by modified gravity is washed away with the 
projection along the line-of-sight. On the other hand, while their prediction for 
\frz$=-10^{-5}$ of a 10 per cent increase in \dksz\ is consistent with our findings 
(magenta line in the bottom panel of Fig.~\ref{f:pows}), albeit slightly smaller, 
for \frz$=-10^{-4}$ they predict a boost in \dksz\ by 30 per cent against our 50 per 
cent (red line). This 
discrepancy is likely an indication that \cite{bianchini16} are underestimating 
the non-linear contribution of the low-$z$ LSS to the kSZ effect. This is also the 
possible explanation of their lower value of \dkszt\ for the GR model. In detail, 
they predict  \dkszt$\simeq3.85\,$\muk\sqr, with cosmological parameters almost identical 
to ours, except for the assumption \zre=9.9 instead of our more conservative \zre=8.8. 
When adopting the same value, our prediction rises to \dkszt$=4.2\,$\muk\sqr\ 
(\zre$=9.9$, volume corrected), thus about 10 per cent higher.

Regarding this point, in a recent work \cite{park16} provided a detailed 
study on the importance of including the connected four-point term in the transverse 
momentum power spectrum \citep[see also][]{park18}. This term, that is usually neglected 
in analytical predictions, becomes non-negligible at wave numbers 
$k\gtrsim1\,h\,$Mpc\mone\ at low redshift, enough to contribute to about 10 per cent the 
of the kSZ effect signal of the LSS computed with the unconnected terms alone. Our 
numerical approach considers both the linear and non-linear regime at high precision, 
therefore confirming the results of \cite{park16}. We stress that this aspect is crucial 
when comparing kSZ power spectrum predictions with observational data, making our results 
potentially interesting to provide constraints on both cosmology and reionization models 
(see also the discussion in Paper~I).

%%%%%%%%%%%%%%%%%%%%%%%%%%%%%%%%%%%%%%%%%%%%%%%%%%%%%%%%%%%%%%%%%%%%%%%%%%%%%%%%%%%%%%%%%%
\subsection{Implications for patchy reionization scenarios} %%%%%%%%%%%%%%%%%%%%%%%%%%%%%%
\label{ss:impl}

Since we are considering only the contribution of the ionised IGM after the EoR, if the 
reionization process in non-homogeneous then the presence of ionized bubbles would 
produce an additional kSZ effect adding to the total value of \dksz. Being the expected 
size of these ``patches'' of the order of $\sim$10 Mpc (comoving), this would contribute 
in particular at $\ell=3000$, i.e. at the angular scale measured by SPT.

By combining our results for the GR simulation with the SPT measurement, we obtain an 
upper limit on the additional kSZ power associated to patchy reionization of 
\dksztp$<0.9\,$\muk\sqr\ (95 per cent C.L.), valid under the assumption of \zre=8.8 and 
with the cosmological parameters quoted in Table~\ref{t:sim}. This limit is very 
stringent and favours homogeneous and/or fast reionization scenarios. Unfortunately, only 
a few works on reionization models provide forecasts for the corresponding contribution 
to the kSZ power spectrum. Among those, the models described in \cite{iliev07} predict 
values in the range \dksztp$\sim2.2-3.7\,$\muk\sqr, depending on the photon production 
efficiency of the ionizing sources, well above the upper limits reported here. The same 
applies to the models of \cite{mesinger12}, with predictions of 
\dksztp$=1.5-3.5\,$\muk\sqr. Moreover, most of the models described in \cite{mesinger13}, 
with \dksztp$=1.2-2.3\,$\muk\sqr, are also ruled out. Only their model model that assumes 
a large contribution of X-ray ionizing sources indicates a significantly smaller 
\dksztp$=0.95\,$\muk\sqr, thus close to the upper limit presented in this work. On the 
other hand, the reionization models by \cite{park13} predict \dksztp$=0.66-0.83\,$
\muk\sqr: these more conservative forecasts are below our upper limit, albeit very close.

In addition, we stress that when including $f(R)$ gravity the corresponding upper 
limits on \dksztp shrink to $<0.7\,$ and $<0.4\,$\muk\sqr\ (95 per cent C.L.) for 
\frz$=10^{-6}$ and $10^{-5}$, respectively. With \frz$=10^{-4}$ \dksztp\ is consistent 
with zero at 99.5 per cent.

%%%%%%%%%%%%%%%%%%%%%%%%%%%%%%%%%%%%%%%%%%%%%%%%%%%%%%%%%%%%%%%%%%%%%%%%%%%%%%%%%%%%%%%%%%
%%%%%%%%%%%%%%%%%%%%%%%%%%%%%%%%% SUMMARY AND CONCLUSIONS %%%%%%%%%%%%%%%%%%%%%%%%%%%%%%%%
%%%%%%%%%%%%%%%%%%%%%%%%%%%%%%%%%%%%%%%%%%%%%%%%%%%%%%%%%%%%%%%%%%%%%%%%%%%%%%%%%%%%%%%%%%

\section{SUMMARY AND CONCLUSIONS} \label{s:concl}

In this paper we presented the first prediction of the properties of the kSZ effect of 
the post-reionization LSS derived from a set of hydrodynamical simulations that include 
the effect of $f(R)$ modified gravity. Using \mggadg\ 
\citep{Puchwein_Baldi_Springel_2013}, we run a set of 4 simulations in a 
$L_{\rm box}=240$\hmone\ Mpc comoving box with the same cosmology but 
with different modified gravity parameters, namely \frzt, where \frz\ is defined 
according to the \cite{Hu_Sawicki_2007} formalism and \frz$=0$ corresponds to the 
standard GR case. Starting from the simulation outputs, we compute a set of 50 
Doppler \bpar\ maps derived from different light-cone realizations, that we use to 
predict the properties of the kSZ effect signal, most notably its power 
spectrum, and quantify how the \frz\ parameter affects them. By keeping track of the 
line-of-sight information, we are able to provide the redshift distribution of the kSZ 
signal, allowing us to study the dependence with \zre, the redshift at 
which reionization occurs.

As in \citet[][our previous study on the effect of massive neutrinos, i.e. Paper~I]
{roncarelli17}  we correct for the missing velocity power due to the limited simulation 
volume with a calibrated analytical formula. We also account for the cosmological star 
mass fraction relying on observational results \citep{ilbert13}. This allows us to 
compare our results on the kSZ effect with the recent SPT measurement of 
\dkszt$=(2.9\pm1.3)\,$\muk\sqr\ by \cite{george15} and to provide stringent upper 
limits on the possible contribution to the kSZ power by patchy reionization.

Our main findings can be summarised as follows.
\begin{enumerate}
\item As in Paper~I, we obtain that the distribution of the Doppler-\bpar\ can be 
described by a Gaussian curve, centered in 0. Its dispersion increases for larger 
absolute values of \frz, scaling as $\sigma_b \propto 
\left(1 +\sqrt{\left|\overline{f}_{\rm R,0}\right|}\right)^{12.5}$.
\item The amplitude of the kSZ power spectrum is boosted by the presence of modified 
gravity. The increase with respect to the GR scenario is about 2--3 per cent for 
\frz$=10^{-6}$ and rises up to 50 per cent for \frz$=10^{-4}$, almost independent on the 
angular scale.  
\item For our fiducial GR model we predict an amplitude of the kSZ power spectrum due to 
the ionized LSS at $\ell=3000$ of \dkszt$=4.1\,$\muk\sqr (\zre$=8.8$), consistent 
with our findings of Paper~I. By combining it with SPT results \citep{george15}, we 
confirm a stringent upper limit to the kSZ contribution due to patchy reionization of 
\dksztp$<0.9\,$\muk\sqr\ (95 per cent C.~L.), thus favouring homogeneous and fast 
reionization scenarios and ruling out many of the current models of the EoR history.
\item In the presence of modified gravity consistent with current constraints, these 
limits shrink to \dksztp$<0.7\,$ for \frz$=10^{-6}$, and $<0.4\,$\muk\sqr\ (95 per cent 
C.L.) for $10^{-5}$.
\item Finally, we studied the scaling of the kSZ power spectrum with \zre\ and \frz\ 
and provide a fitting formula, equation~(\ref{e:fit}), that works for 
$2000<\ell<20000$. The best-fit parameters are shown in Table~\ref{t:fit}. The scaling 
with \frz\ is approximately $\mathcal{D}^{\rm kSZ}_{\ell} \propto 
\left(1+\sqrt{\left|\overline{f}_{\rm R,0}\right|}\right)^{42}$.
\end{enumerate}

Our work confirms that the kSZ effect of the LSS is an extremely promising probe of the 
high redshift Universe. In fact, despite the little observational data available, with 
an accurate modelling of the LSS after the EoR it is already possible to obtain 
interesting constraints on both non-standard cosmology and reionization models. We also 
show the importance of hydrodynamical simulations that prove to be competitive with 
analytical estimates. 

In the future we plan to extend our work with other non-standard cosmological models, 
such as quintessence and coupled dark energy. Since these scenarios are expected to 
enhance the amplitude of the kSZ power spectrum, this will allow to obtain further 
constraints on the nature of dark energy, complementary with other probes, as well as 
studying the degeneracies with the different cosmological parameters beyond the \lcdm\ 
model.

%%%%%%%%%%%%%%%%%%%%%%%%%%%%%%%%%%%%%%%%%%%%%%%%%%%%%%%%%%%%%%%%%%%%%%%%%%%%%%%%%%%%%%%%%%
\section*{Acknowledgements} %%%%%%%%%%%%%%%%%%%%%%%%%%%%%%%%%%%%%%%%%%%%%%%%%%%%%%%%%%%%%%

This work has been done despite the shameful situation of the Italian research system, 
worsened by an entire decade of severe cuts to the fundings of public universities and 
research insititutes \cite[see, e.g.,][]{abbott06,abbott16,abbott18}. With some 
noticeable -- yet statistically negligible -- exceptions, this has caused a 
whole generation of valuable researchers, in all fields, to struggle in poor working 
conditions with little hope to achieve decent employment contracts and permanent 
positions, resulting in obvious difficulties in the planning of future research 
activities and in open violation of the European Charter for 
Researchers\footnote{\href{https://euraxess.ec.europa.eu/jobs/charter}
{https://euraxess.ec.europa.eu/jobs/charter}}. 

This work has been supported by ASI (Italian Space Agency) through the Contract 
n.~2015-046-R.0. MR also acknowledges financial contribution from the agreement ASI 
n.~I/023/12/0 ``\emph{Attivit\`a relative alla fase B2/C per la missione Euclid}''. 
The computational analyses have been done thanks to CPU time granted by the ISCRA-CINECA
project ``\emph{The kinetic SZ effect as a cosmological probe}''. The 
work of FVN is supported by the Simons Foundation. MB acknowledges support from the 
Italian Ministry for Education, University and Research (MIUR) through the SIR individual 
grant SIMCODE, project number RBSI14P4IH. We also thank O.~Cucciati, L.~Moscardini, 
E.~Vanzella, M.~Viel and G.~Zamorani for useful suggestions and discussions. We are 
grateful to F.~Bianchini and A.~Silvestri for reading the draft and providing useful 
comments.

\bibliographystyle{mnras}

\begin{thebibliography}{}
\makeatletter
\relax
\def\mn@urlcharsother{\let\do\@makeother \do\$\do\&\do\#\do\^\do\_\do\%\do\~}
\def\mn@doi{\begingroup\mn@urlcharsother \@ifnextchar [ {\mn@doi@}
  {\mn@doi@[]}}
\def\mn@doi@[#1]#2{\def\@tempa{#1}\ifx\@tempa\@empty \href
  {http://dx.doi.org/#2} {doi:#2}\else \href {http://dx.doi.org/#2} {#1}\fi
  \endgroup}
\def\mn@eprint#1#2{\mn@eprint@#1:#2::\@nil}
\def\mn@eprint@arXiv#1{\href {http://arxiv.org/abs/#1} {{\tt arXiv:#1}}}
\def\mn@eprint@dblp#1{\href {http://dblp.uni-trier.de/rec/bibtex/#1.xml}
  {dblp:#1}}
\def\mn@eprint@#1:#2:#3:#4\@nil{\def\@tempa {#1}\def\@tempb {#2}\def\@tempc
  {#3}\ifx \@tempc \@empty \let \@tempc \@tempb \let \@tempb \@tempa \fi \ifx
  \@tempb \@empty \def\@tempb {arXiv}\fi \@ifundefined
  {mn@eprint@\@tempb}{\@tempb:\@tempc}{\expandafter \expandafter \csname
  mn@eprint@\@tempb\endcsname \expandafter{\@tempc}}}

\bibitem[\protect\citeauthoryear{{Abbott}}{{Abbott}}{2006}]{abbott06}
{Abbott} A.,  2006, \mn@doi [\nat] {10.1038/440264a}, \href
  {http://adsabs.harvard.edu/abs/2006Natur.440..264A} {440, 264}

\bibitem[\protect\citeauthoryear{{Abbott}}{{Abbott}}{2016}]{abbott16}
{Abbott} A.,  2016, \mn@doi [\nat] {10.1038/nature.2016.21139}, \href
  {http://adsabs.harvard.edu/abs/2016Natur.540..324A} {540, 324}

\bibitem[\protect\citeauthoryear{{Abbott}}{{Abbott}}{2018}]{abbott18}
{Abbott} A.,  2018, \mn@doi [\nat] {10.1038/d41586-018-02223-7}, 554, 411

\bibitem[\protect\citeauthoryear{Abbott et~al.,}{Abbott
  et~al.}{2017}]{GW170817}
Abbott B.~P.,  et~al., 2017, \mn@doi [Phys. Rev. Lett.]
  {10.1103/PhysRevLett.119.161101}, 119, 161101

\bibitem[\protect\citeauthoryear{{Amendola}}{{Amendola}}{2000}]{Amendola_2000}
{Amendola} L.,  2000, \mn@doi [\prd] {10.1103/PhysRevD.62.043511}, \href
  {http://adsabs.harvard.edu/abs/2000PhRvD..62d3511A} {62, 043511}

\bibitem[\protect\citeauthoryear{{Amendola}, {Baldi}  \&
  {Wetterich}}{{Amendola} et~al.}{2008}]{Amendola_Baldi_Wetterich_2008}
{Amendola} L.,  {Baldi} M.,   {Wetterich} C.,  2008, \mn@doi [\prd]
  {10.1103/PhysRevD.78.023015}, \href
  {http://adsabs.harvard.edu/abs/2008PhRvD..78b3015A} {78, 023015}

\bibitem[\protect\citeauthoryear{{Amendola} et~al.}{{Amendola}
  et~al.}{2013}]{Euclid_TWG}
{Amendola} L.,  et~al., 2013, \mn@doi [Living Reviews in Relativity]
  {10.12942/lrr-2013-6}, \href
  {http://adsabs.harvard.edu/abs/2013LRR....16....6A} {16, 6}

\bibitem[\protect\citeauthoryear{{Baker}, {Bellini}, {Ferreira}, {Lagos},
  {Noller}  \& {Sawicki}}{{Baker} et~al.}{2017}]{Baker_etal_2017}
{Baker} T.,  {Bellini} E.,  {Ferreira} P.~G.,  {Lagos} M.,  {Noller} J.,
  {Sawicki} I.,  2017, \mn@doi [Physical Review Letters]
  {10.1103/PhysRevLett.119.251301}, \href
  {http://adsabs.harvard.edu/abs/2017PhRvL.119y1301B} {119, 251301}

\bibitem[\protect\citeauthoryear{{Baldi}, {Villaescusa-Navarro}, {Viel},
  {Puchwein}, {Springel}  \& {Moscardini}}{{Baldi} et~al.}{2014}]{baldi14}
{Baldi} M.,  {Villaescusa-Navarro} F.,  {Viel} M.,  {Puchwein} E.,  {Springel}
  V.,   {Moscardini} L.,  2014, \mn@doi [\mnras] {10.1093/mnras/stu259}, \href
  {http://adsabs.harvard.edu/abs/2014MNRAS.440...75B} {440, 75}

\bibitem[\protect\citeauthoryear{{Battaglia}, {Bond}, {Pfrommer}, {Sievers}  \&
  {Sijacki}}{{Battaglia} et~al.}{2010}]{battaglia10}
{Battaglia} N.,  {Bond} J.~R.,  {Pfrommer} C.,  {Sievers} J.~L.,   {Sijacki}
  D.,  2010, \mn@doi [\apj] {10.1088/0004-637X/725/1/91}, \href
  {http://adsabs.harvard.edu/abs/2010ApJ...725...91B} {725, 91}

\bibitem[\protect\citeauthoryear{{Bertotti}, {Iess}  \& {Tortora}}{{Bertotti}
  et~al.}{2003}]{Bertotti_Iess_Tortora_2003}
{Bertotti} B.,  {Iess} L.,   {Tortora} P.,  2003, \mn@doi [\nat]
  {10.1038/nature01997}, \href
  {http://adsabs.harvard.edu/abs/2003Natur.425..374B} {425, 374}

\bibitem[\protect\citeauthoryear{{Bianchini} \& {Silvestri}}{{Bianchini} \&
  {Silvestri}}{2016}]{bianchini16}
{Bianchini} F.,  {Silvestri} A.,  2016, \mn@doi [\prd]
  {10.1103/PhysRevD.93.064026}, \href
  {http://adsabs.harvard.edu/abs/2016PhRvD..93f4026B} {93, 064026}

\bibitem[\protect\citeauthoryear{{Buchdahl}}{{Buchdahl}}{1970}]{Buchdahl_1970}
{Buchdahl} H.~A.,  1970, \mn@doi [\mnras] {10.1093/mnras/150.1.1}, \href
  {http://adsabs.harvard.edu/abs/1970MNRAS.150....1B} {150, 1}

\bibitem[\protect\citeauthoryear{{Crawford} et~al.}{{Crawford}
  et~al.}{2014}]{crawford14}
{Crawford} T.~M.,  et~al., 2014, \mn@doi [\apj] {10.1088/0004-637X/784/2/143},
  \href {http://adsabs.harvard.edu/abs/2014ApJ...784..143C} {784, 143}

\bibitem[\protect\citeauthoryear{{Creminelli} \& {Vernizzi}}{{Creminelli} \&
  {Vernizzi}}{2017}]{Creminelli_Vernizzi_2017}
{Creminelli} P.,  {Vernizzi} F.,  2017, \mn@doi [Physical Review Letters]
  {10.1103/PhysRevLett.119.251302}, \href
  {http://adsabs.harvard.edu/abs/2017PhRvL.119y1302C} {119, 251302}

\bibitem[\protect\citeauthoryear{{Damour} \& {Polyakov}}{{Damour} \&
  {Polyakov}}{1994}]{Damour_Polyakov_1994}
{Damour} T.,  {Polyakov} A.~M.,  1994, \mn@doi [Nuclear Physics B]
  {10.1016/0550-3213(94)90143-0}, \href
  {http://adsabs.harvard.edu/abs/1994NuPhB.423..532D} {423, 532}

\bibitem[\protect\citeauthoryear{{De Felice} \& {Tsujikawa}}{{De Felice} \&
  {Tsujikawa}}{2010}]{DeFelice_Tsujikawa_2010}
{De Felice} A.,  {Tsujikawa} S.,  2010, \mn@doi [Living Reviews in Relativity]
  {10.12942/lrr-2010-3}, \href
  {http://adsabs.harvard.edu/abs/2010LRR....13....3D} {13, 3}

\bibitem[\protect\citeauthoryear{{George} et~al.}{{George}
  et~al.}{2015}]{george15}
{George} E.~M.,  et~al., 2015, \mn@doi [\apj] {10.1088/0004-637X/799/2/177},
  \href {http://adsabs.harvard.edu/abs/2015ApJ...799..177G} {799, 177}

\bibitem[\protect\citeauthoryear{{He}}{{He}}{2013}]{He_2013}
{He} J.-h.,  2013, arXiv:1307.4876, \href
  {http://adsabs.harvard.edu/abs/2013arXiv1307.4876H} {}

\bibitem[\protect\citeauthoryear{{Hinterbichler} \& {Khoury}}{{Hinterbichler}
  \& {Khoury}}{2010}]{Hinterbichler_Khoury_2010}
{Hinterbichler} K.,  {Khoury} J.,  2010, \mn@doi [Physical Review Letters]
  {10.1103/PhysRevLett.104.231301}, \href
  {http://adsabs.harvard.edu/abs/2010PhRvL.104w1301H} {104, 231301}

\bibitem[\protect\citeauthoryear{{Hu} \& {Sawicki}}{{Hu} \&
  {Sawicki}}{2007}]{Hu_Sawicki_2007}
{Hu} W.,  {Sawicki} I.,  2007, \mn@doi [\prd] {10.1103/PhysRevD.76.064004},
  \href {http://adsabs.harvard.edu/abs/2007PhRvD..76f4004H} {76, 064004}

\bibitem[\protect\citeauthoryear{{Hu}, {Raveri}, {Rizzato}  \&
  {Silvestri}}{{Hu} et~al.}{2016}]{hu16}
{Hu} B.,  {Raveri} M.,  {Rizzato} M.,   {Silvestri} A.,  2016, \mn@doi [\mnras]
  {10.1093/mnras/stw775}, \href
  {http://adsabs.harvard.edu/abs/2016MNRAS.459.3880H} {459, 3880}

\bibitem[\protect\citeauthoryear{{Ilbert} et~al.}{{Ilbert}
  et~al.}{2013}]{ilbert13}
{Ilbert} O.,  et~al., 2013, \mn@doi [\aap] {10.1051/0004-6361/201321100}, \href
  {http://adsabs.harvard.edu/abs/2013A%26A...556A..55I} {556, A55}

\bibitem[\protect\citeauthoryear{{Iliev}, {Pen}, {Bond}, {Mellema}  \&
  {Shapiro}}{{Iliev} et~al.}{2007}]{iliev07}
{Iliev} I.~T.,  {Pen} U.-L.,  {Bond} J.~R.,  {Mellema} G.,   {Shapiro} P.~R.,
  2007, \mn@doi [\apj] {10.1086/513687}, \href
  {http://adsabs.harvard.edu/abs/2007ApJ...660..933I} {660, 933}

\bibitem[\protect\citeauthoryear{{Iliev}, {Mellema}, {Ahn}, {Shapiro}, {Mao}
  \& {Pen}}{{Iliev} et~al.}{2014}]{iliev14}
{Iliev} I.~T.,  {Mellema} G.,  {Ahn} K.,  {Shapiro} P.~R.,  {Mao} Y.,   {Pen}
  U.-L.,  2014, \mn@doi [\mnras] {10.1093/mnras/stt2497}, \href
  {http://adsabs.harvard.edu/abs/2014MNRAS.439..725I} {439, 725}

\bibitem[\protect\citeauthoryear{{Jain} \& {VanderPlas}}{{Jain} \&
  {VanderPlas}}{2011}]{Jain_VanderPlas_2011}
{Jain} B.,  {VanderPlas} J.,  2011, \mn@doi [\jcap]
  {10.1088/1475-7516/2011/10/032}, \href
  {http://adsabs.harvard.edu/abs/2011JCAP...10..032J} {10, 032}

\bibitem[\protect\citeauthoryear{{Jain}, {Vikram}  \& {Sakstein}}{{Jain}
  et~al.}{2013}]{Jain_Vikram_Sakstein_2013}
{Jain} B.,  {Vikram} V.,   {Sakstein} J.,  2013, \mn@doi [\apj]
  {10.1088/0004-637X/779/1/39}, \href
  {http://adsabs.harvard.edu/abs/2013ApJ...779...39J} {779, 39}

\bibitem[\protect\citeauthoryear{{Khoury} \& {Weltman}}{{Khoury} \&
  {Weltman}}{2004}]{Khoury_Weltman_2004}
{Khoury} J.,  {Weltman} A.,  2004, \mn@doi [\prd] {10.1103/PhysRevD.69.044026},
  \href {http://adsabs.harvard.edu/abs/2004PhRvD..69d4026K} {69, 044026}

\bibitem[\protect\citeauthoryear{{Laureijs} et~al.,}{{Laureijs}
  et~al.}{2011}]{Euclid-r}
{Laureijs} R.,  et~al., 2011, preprint, \href
  {http://adsabs.harvard.edu/abs/2011arXiv1110.3193L} {} (\mn@eprint {arXiv}
  {1110.3193})

\bibitem[\protect\citeauthoryear{{Lewis}, {Challinor}  \& {Lasenby}}{{Lewis}
  et~al.}{2000}]{CAMB}
{Lewis} A.,  {Challinor} A.,   {Lasenby} A.,  2000, \mn@doi [\apj]
  {10.1086/309179}, \href {http://adsabs.harvard.edu/abs/2000ApJ...538..473L}
  {538, 473}

\bibitem[\protect\citeauthoryear{{Lombriser}}{{Lombriser}}{2014}]{Lombriser_2014}
{Lombriser} L.,  2014, \mn@doi [Annalen der Physik] {10.1002/andp.201400058},
  \href {http://adsabs.harvard.edu/abs/2014AnP...526..259L} {526, 259}

\bibitem[\protect\citeauthoryear{{Lombriser}, {Koyama}  \& {Li}}{{Lombriser}
  et~al.}{2014}]{Lombriser_Koyama_Li_2014}
{Lombriser} L.,  {Koyama} K.,   {Li} B.,  2014, \mn@doi [\jcap]
  {10.1088/1475-7516/2014/03/021}, \href
  {http://adsabs.harvard.edu/abs/2014JCAP...03..021L} {3, 021}

\bibitem[\protect\citeauthoryear{{Mar{\'{\i}}a Ezquiaga} \&
  {Zumalac{\'a}rregui}}{{Mar{\'{\i}}a Ezquiaga} \&
  {Zumalac{\'a}rregui}}{2017}]{Ezquiaga_Zumalacarregui_2017}
{Mar{\'{\i}}a Ezquiaga} J.,  {Zumalac{\'a}rregui} M.,  2017, preprint, \href
  {http://adsabs.harvard.edu/abs/2017arXiv171005901M} {} (\mn@eprint {arXiv}
  {1710.05901})

\bibitem[\protect\citeauthoryear{{Mesinger}, {McQuinn}  \&
  {Spergel}}{{Mesinger} et~al.}{2012}]{mesinger12}
{Mesinger} A.,  {McQuinn} M.,   {Spergel} D.~N.,  2012, \mn@doi [\mnras]
  {10.1111/j.1365-2966.2012.20713.x}, \href
  {http://adsabs.harvard.edu/abs/2012MNRAS.422.1403M} {422, 1403}

\bibitem[\protect\citeauthoryear{{Mesinger}, {Ferrara}  \&
  {Spiegel}}{{Mesinger} et~al.}{2013}]{mesinger13}
{Mesinger} A.,  {Ferrara} A.,   {Spiegel} D.~S.,  2013, \mn@doi [\mnras]
  {10.1093/mnras/stt198}, \href
  {http://adsabs.harvard.edu/abs/2013MNRAS.431..621M} {431, 621}

\bibitem[\protect\citeauthoryear{Motohashi, Starobinsky  \& Yokoyama}{Motohashi
  et~al.}{2013}]{Motohashi_etal_2013}
Motohashi H.,  Starobinsky A.~A.,   Yokoyama J.,  2013, \mn@doi
  [Phys.Rev.Lett.] {10.1103/PhysRevLett.110.121302}, 110, 121302

\bibitem[\protect\citeauthoryear{{Nicolis}, {Rattazzi}  \&
  {Trincherini}}{{Nicolis} et~al.}{2009}]{Nicolis_Rattazzi_Trincherini_2009}
{Nicolis} A.,  {Rattazzi} R.,   {Trincherini} E.,  2009, \mn@doi [\prd]
  {10.1103/PhysRevD.79.064036}, \href
  {http://adsabs.harvard.edu/abs/2009PhRvD..79f4036N} {79, 064036}

\bibitem[\protect\citeauthoryear{{Ostriker} \& {Vishniac}}{{Ostriker} \&
  {Vishniac}}{1986}]{ostriker86}
{Ostriker} J.~P.,  {Vishniac} E.~T.,  1986, \mn@doi [\apjl] {10.1086/184704},
  \href {http://adsabs.harvard.edu/abs/1986ApJ...306L..51O} {306, L51}

\bibitem[\protect\citeauthoryear{{Park}, {Shapiro}, {Komatsu}, {Iliev}, {Ahn}
  \& {Mellema}}{{Park} et~al.}{2013}]{park13}
{Park} H.,  {Shapiro} P.~R.,  {Komatsu} E.,  {Iliev} I.~T.,  {Ahn} K.,
  {Mellema} G.,  2013, \mn@doi [\apj] {10.1088/0004-637X/769/2/93}, \href
  {http://adsabs.harvard.edu/abs/2013ApJ...769...93P} {769, 93}

\bibitem[\protect\citeauthoryear{{Park}, {Komatsu}, {Shapiro}, {Koda}  \&
  {Mao}}{{Park} et~al.}{2016}]{park16}
{Park} H.,  {Komatsu} E.,  {Shapiro} P.~R.,  {Koda} J.,   {Mao} Y.,  2016,
  \mn@doi [\apj] {10.3847/0004-637X/818/1/37}, \href
  {http://adsabs.harvard.edu/abs/2016ApJ...818...37P} {818, 37}

\bibitem[\protect\citeauthoryear{{Park}, {Alvarez}  \& {Bond}}{{Park}
  et~al.}{2018}]{park18}
{Park} H.,  {Alvarez} M.~A.,   {Bond} J.~R.,  2018, \mn@doi [\apj]
  {10.3847/1538-4357/aaa0da}, \href
  {http://adsabs.harvard.edu/abs/2018ApJ...853..121P} {853, 121}

\bibitem[\protect\citeauthoryear{Peel, Pettorino, Giocoli, Starck  \&
  Baldi}{Peel et~al.}{2018}]{Peel_etal_2018}
Peel A.,  Pettorino V.,  Giocoli C.,  Starck J.-L.,   Baldi M.,  2018, in prep

\bibitem[\protect\citeauthoryear{{Perlmutter} et~al.}{{Perlmutter}
  et~al.}{1999}]{Perlmutter_etal_1999}
{Perlmutter} S.,  et~al., 1999, \mn@doi [\apj] {10.1086/307221}, \href
  {http://adsabs.harvard.edu/abs/1999ApJ...517..565P} {517, 565}

\bibitem[\protect\citeauthoryear{{Planck Collaboration XIII}}{{Planck
  Collaboration XIII}}{2016}]{planck16cp}
{Planck Collaboration XIII} 2016, \mn@doi [\aap] {10.1051/0004-6361/201525830},
  \href {http://adsabs.harvard.edu/abs/2016A%26A...594A..13P} {594, A13}

\bibitem[\protect\citeauthoryear{{Puchwein}, {Baldi}  \& {Springel}}{{Puchwein}
  et~al.}{2013}]{Puchwein_Baldi_Springel_2013}
{Puchwein} E.,  {Baldi} M.,   {Springel} V.,  2013, \mn@doi [\mnras]
  {10.1093/mnras/stt1575}, \href
  {http://adsabs.harvard.edu/abs/2013MNRAS.436..348P} {436, 348}

\bibitem[\protect\citeauthoryear{{Ratra} \& {Peebles}}{{Ratra} \&
  {Peebles}}{1988}]{Ratra_Peebles_1988}
{Ratra} B.,  {Peebles} P.~J.~E.,  1988, \mn@doi [\prd]
  {10.1103/PhysRevD.37.3406}, \href
  {http://adsabs.harvard.edu/abs/1988PhRvD..37.3406R} {37, 3406}

\bibitem[\protect\citeauthoryear{{Riess} et~al.,}{{Riess}
  et~al.}{1998}]{Riess_etal_1998}
{Riess} A.~G.,  et~al., 1998, \mn@doi [\aj] {10.1086/300499}, \href
  {http://adsabs.harvard.edu/abs/1998AJ....116.1009R} {116, 1009}

\bibitem[\protect\citeauthoryear{{Roncarelli}, {Moscardini}, {Borgani}  \&
  {Dolag}}{{Roncarelli} et~al.}{2007}]{roncarelli07}
{Roncarelli} M.,  {Moscardini} L.,  {Borgani} S.,   {Dolag} K.,  2007, \mn@doi
  [\mnras] {10.1111/j.1365-2966.2007.11914.x}, \href
  {http://adsabs.harvard.edu/abs/2007MNRAS.378.1259R} {378, 1259}

\bibitem[\protect\citeauthoryear{{Roncarelli}, {Moscardini}, {Branchini},
  {Dolag}, {Grossi}, {Iannuzzi}  \& {Matarrese}}{{Roncarelli}
  et~al.}{2010}]{roncarelli10a}
{Roncarelli} M.,  {Moscardini} L.,  {Branchini} E.,  {Dolag} K.,  {Grossi} M.,
  {Iannuzzi} F.,   {Matarrese} S.,  2010, \mn@doi [\mnras]
  {10.1111/j.1365-2966.2009.15978.x}, \href
  {http://adsabs.harvard.edu/abs/2010MNRAS.402..923R} {402, 923}

\bibitem[\protect\citeauthoryear{{Roncarelli}, {Cappelluti}, {Borgani},
  {Branchini}  \& {Moscardini}}{{Roncarelli} et~al.}{2012}]{roncarelli12}
{Roncarelli} M.,  {Cappelluti} N.,  {Borgani} S.,  {Branchini} E.,
  {Moscardini} L.,  2012, \mn@doi [\mnras] {10.1111/j.1365-2966.2012.21277.x},
  \href {http://adsabs.harvard.edu/abs/2012MNRAS.424.1012R} {424, 1012}

\bibitem[\protect\citeauthoryear{{Roncarelli}, {Carbone}  \&
  {Moscardini}}{{Roncarelli} et~al.}{2015}]{roncarelli15}
{Roncarelli} M.,  {Carbone} C.,   {Moscardini} L.,  2015, \mn@doi [\mnras]
  {10.1093/mnras/stu2546}, \href
  {http://adsabs.harvard.edu/abs/2015MNRAS.447.1761R} {447, 1761}

\bibitem[\protect\citeauthoryear{{Roncarelli}, {Villaescusa-Navarro}  \&
  {Baldi}}{{Roncarelli} et~al.}{2017}]{roncarelli17}
{Roncarelli} M.,  {Villaescusa-Navarro} F.,   {Baldi} M.,  2017, \mn@doi
  [\mnras] {10.1093/mnras/stx170}, \href
  {http://adsabs.harvard.edu/abs/2017MNRAS.467..985R} {467, 985}

\bibitem[\protect\citeauthoryear{{Roncarelli}\noopsort{a}, {Moscardini},
  {Tozzi}, {Borgani}, {Cheng}, {Diaferio}, {Dolag}  \&
  {Murante}}{{Roncarelli}\noopsort{a} et~al.}{2006}]{roncarelli06a}
{Roncarelli}\noopsort{a} M.,  {Moscardini} L.,  {Tozzi} P.,  {Borgani} S.,
  {Cheng} L.~M.,  {Diaferio} A.,  {Dolag} K.,   {Murante} G.,  2006, \mn@doi
  [\mnras] {10.1111/j.1365-2966.2006.10102.x}, \href
  {http://adsabs.harvard.edu/abs/2006MNRAS.368...74R} {368, 74}

\bibitem[\protect\citeauthoryear{{Sakstein} \& {Jain}}{{Sakstein} \&
  {Jain}}{2017}]{Sakstein_Jain_2017}
{Sakstein} J.,  {Jain} B.,  2017, \mn@doi [Physical Review Letters]
  {10.1103/PhysRevLett.119.251303}, \href
  {http://adsabs.harvard.edu/abs/2017PhRvL.119y1303S} {119, 251303}

\bibitem[\protect\citeauthoryear{{Schmidt} et~al.,}{{Schmidt}
  et~al.}{1998}]{Schmidt_etal_1998}
{Schmidt} B.~P.,  et~al., 1998, \mn@doi [\apj] {10.1086/306308}, \href
  {http://adsabs.harvard.edu/abs/1998ApJ...507...46S} {507, 46}

\bibitem[\protect\citeauthoryear{{Shaw}, {Rudd}  \& {Nagai}}{{Shaw}
  et~al.}{2012}]{shaw12}
{Shaw} L.~D.,  {Rudd} D.~H.,   {Nagai} D.,  2012, \mn@doi [\apj]
  {10.1088/0004-637X/756/1/15}, \href
  {http://adsabs.harvard.edu/abs/2012ApJ...756...15S} {756, 15}

\bibitem[\protect\citeauthoryear{{Sotiriou} \& {Faraoni}}{{Sotiriou} \&
  {Faraoni}}{2010}]{Sotiriou_Faraoni_2010}
{Sotiriou} T.~P.,  {Faraoni} V.,  2010, \mn@doi [Reviews of Modern Physics]
  {10.1103/RevModPhys.82.451}, \href
  {http://adsabs.harvard.edu/abs/2010RvMP...82..451S} {82, 451}

\bibitem[\protect\citeauthoryear{{Springel}}{{Springel}}{2005}]{springel05}
{Springel} V.,  2005, \mn@doi [\mnras] {10.1111/j.1365-2966.2005.09655.x},
  \href {http://adsabs.harvard.edu/abs/2005MNRAS.364.1105S} {364, 1105}

\bibitem[\protect\citeauthoryear{{Sunyaev} \& {Zeldovich}}{{Sunyaev} \&
  {Zeldovich}}{1970}]{sunyaev70}
{Sunyaev} R.~A.,  {Zeldovich} Y.~B.,  1970, \mn@doi [\apss]
  {10.1007/BF00653471}, \href
  {http://adsabs.harvard.edu/abs/1970Ap%26SS...7....3S} {7, 3}

\bibitem[\protect\citeauthoryear{{Trac}, {Bode}  \& {Ostriker}}{{Trac}
  et~al.}{2011}]{trac11}
{Trac} H.,  {Bode} P.,   {Ostriker} J.~P.,  2011, \mn@doi [\apj]
  {10.1088/0004-637X/727/2/94}, \href
  {http://adsabs.harvard.edu/abs/2011ApJ...727...94T} {727, 94}

\bibitem[\protect\citeauthoryear{{Ursino}, {Galeazzi}  \&
  {Roncarelli}}{{Ursino} et~al.}{2010}]{ursino10}
{Ursino} E.,  {Galeazzi} M.,   {Roncarelli} M.,  2010, \mn@doi [\apj]
  {10.1088/0004-637X/721/1/46}, \href
  {http://adsabs.harvard.edu/abs/2010ApJ...721...46U} {721, 46}

\bibitem[\protect\citeauthoryear{Vainshtein}{Vainshtein}{1972}]{Vainshtein_1972}
Vainshtein A.,  1972, \mn@doi [Phys.Lett.] {10.1016/0370-2693(72)90147-5}, B39,
  393

\bibitem[\protect\citeauthoryear{{Viel}, {Lesgourgues}, {Haehnelt}, {Matarrese}
   \& {Riotto}}{{Viel} et~al.}{2005}]{Viel_etal_2005}
{Viel} M.,  {Lesgourgues} J.,  {Haehnelt} M.~G.,  {Matarrese} S.,   {Riotto}
  A.,  2005, \mn@doi [\prd] {10.1103/PhysRevD.71.063534}, \href
  {http://adsabs.harvard.edu/abs/2005PhRvD..71f3534V} {71, 063534}

\bibitem[\protect\citeauthoryear{{Vikram}, {Cabr{\'e}}, {Jain}  \&
  {VanderPlas}}{{Vikram} et~al.}{2013}]{Vikram_etal_2013}
{Vikram} V.,  {Cabr{\'e}} A.,  {Jain} B.,   {VanderPlas} J.~T.,  2013, \mn@doi
  [\jcap] {10.1088/1475-7516/2013/08/020}, \href
  {http://adsabs.harvard.edu/abs/2013JCAP...08..020V} {8, 020}

\bibitem[\protect\citeauthoryear{{Vishniac}}{{Vishniac}}{1987}]{vishniac87}
{Vishniac} E.~T.,  1987, \mn@doi [\apj] {10.1086/165755}, \href
  {http://adsabs.harvard.edu/abs/1987ApJ...322..597V} {322, 597}

\bibitem[\protect\citeauthoryear{{Wetterich}}{{Wetterich}}{1988}]{Wetterich_1988}
{Wetterich} C.,  1988, \mn@doi [Nuclear Physics B]
  {10.1016/0550-3213(88)90193-9}, \href
  {http://adsabs.harvard.edu/abs/1988NuPhB.302..668W} {302, 668}

\bibitem[\protect\citeauthoryear{{Wetterich}}{{Wetterich}}{1995}]{Wetterich_1995}
{Wetterich} C.,  1995, \aap, \href
  {http://adsabs.harvard.edu/abs/1995A%26A...301..321W} {301, 321}

\bibitem[\protect\citeauthoryear{{Will}}{{Will}}{2006}]{Will_2005}
{Will} C.~M.,  2006, \mn@doi [Living Reviews in Relativity]
  {10.12942/lrr-2006-3}, \href
  {http://adsabs.harvard.edu/abs/2006LRR.....9....3W} {9, 3}

\bibitem[\protect\citeauthoryear{{Winther} et~al.,}{{Winther}
  et~al.}{2015}]{Winther_etal_2015}
{Winther} H.~A.,  et~al., 2015, \mn@doi [\mnras] {10.1093/mnras/stv2253}, \href
  {http://adsabs.harvard.edu/abs/2015MNRAS.454.4208W} {454, 4208}

\bibitem[\protect\citeauthoryear{{Wright}, {Winther}  \& {Koyama}}{{Wright}
  et~al.}{2017}]{Wright_Winther_Koyama_2017}
{Wright} B.~S.,  {Winther} H.~A.,   {Koyama} K.,  2017, \mn@doi [\jcap]
  {10.1088/1475-7516/2017/10/054}, \href
  {http://adsabs.harvard.edu/abs/2017JCAP...10..054W} {10, 054}

\bibitem[\protect\citeauthoryear{{Zel'dovich}}{{Zel'dovich}}{1970}]{Zeldovich_1970}
{Zel'dovich} Y.~B.,  1970, \aap, \href
  {http://adsabs.harvard.edu/abs/1970A%26A.....5...84Z} {5, 84}

\bibitem[\protect\citeauthoryear{{Zhao}}{{Zhao}}{2014}]{zhao14}
{Zhao} G.-B.,  2014, \mn@doi [\apjs] {10.1088/0067-0049/211/2/23}, \href
  {http://adsabs.harvard.edu/abs/2014ApJS..211...23Z} {211, 23}

\bibitem[\protect\citeauthoryear{{Zhao}, {Pogosian}, {Silvestri}  \&
  {Zylberberg}}{{Zhao} et~al.}{2009}]{zhao09}
{Zhao} G.-B.,  {Pogosian} L.,  {Silvestri} A.,   {Zylberberg} J.,  2009,
  \mn@doi [\prd] {10.1103/PhysRevD.79.083513}, \href
  {http://adsabs.harvard.edu/abs/2009PhRvD..79h3513Z} {79, 083513}

\makeatother
\end{thebibliography}
\newcommand{\noopsort}[1]{}

\label{lastpage}
\end{document}